\documentclass[prd,onecolumn]{revtex4}
\usepackage{dcolumn}
\usepackage{multirow}
\usepackage{graphicx}
\usepackage{amssymb}
\usepackage{bm}
\usepackage{hyperref}%for .pdf
\usepackage{epstopdf}%for .pdf
\usepackage{color}
\usepackage{mathrsfs}
\usepackage{amsmath,amssymb,amsthm}
\begin{document}
\title{Improved constraints on the dark energy equation of state using Gaussian processes}
\author{Deng Wang}
\email{Cstar@mail.nankai.edu.cn}
\affiliation{Theoretical Physics Division, Chern Institute of Mathematics, Nankai University,
Tianjin 300071, China}
\author{Xin-He Meng}
\email{xhm@nankai.edu.cn}
\affiliation{{Department of Physics, Nankai University, Tianjin 300071, China}}
\begin{abstract}
We perform a comprehensive study of the dark energy equation of state (EoS) utilizing the model-independent Gaussian processes (GP). Using a combination of the Union 2.1 data set, the 30 newly added H(z) cosmic chronometer data points and Planck's shift parameter, we modify the usual GaPP code and provide a tighter constraint on the dark energy EoS than the previous literature about GP reconstructions. Subsequently, we take the `` controlling variable method '' to  investigate directly the effects of variable matter density parameter $\Omega_{m0}$, variable cosmic curvature $\Omega_{k0}$ and variable Hubble constant $H_0$ on the dark energy EoS, respectively. We find that too small or large $\Omega_{m0}$, $\Omega_{k0}$ and $H_0$ are all disfavored by our GP reconstructions based on current cosmological observations. Subsequently, we find that variable $\Omega_{m0}$ and $\Omega_{k0}$ affect the reconstructions of the dark energy EoS, but affect hardly the reconstructions of the normalized comoving distance $D(z)$ and its derivatives $D'(z)$ and $D''(z)$. However, variable $H_0$ affects the reconstructions of the dark energy EoS by affecting obviously those of $D(z), D'(z)$ and $D''(z)$. Furthermore, we find that the results of our reconstructions support substantially the recent local measurement of $H_0$ reported by Riess et al.

\end{abstract}
\maketitle
\section{Introduction}
The elegant discovery that our universe is in accelerating expansion \cite{1,2}, has inspired a large number of studies to explore the cosmological origin and nature of the current amazing phenomena. Due to a lack of deeper understanding at present, cosmologists have introduced an exotic energy component named dark energy to explain the cause of acceleration. As is well known, the simplest candidate of dark energy is the so-called cosmological constant, namely the $\Lambda$-cold-dark-matter ($\Lambda$CDM) model, which is proved to be substantially successful in describing many aspects of the observational universe. For instance, the large scale structure (LSS) of matter distribution at the linear level, the spectrum of anisotropies of the cosmological microwave background (CMB) radiation, and the expansion phenomena are well described by the $\Lambda$CDM model. Nonetheless, this model has already faced two fatal problems, i.e., `` coincidence '' problem and the `` fine-tuning '' problem \cite{3}. The former implies why the dark matter and dark energy are at the same order at present since their densities are so different during the evolution of the universe, while the latter suggests that theoretical estimates for the vacuum density are many orders of magnitude larger than its observed value, namely, the famous 120-orders-of-magnitude discrepancy that makes the vacuum explanation incredulous. This implies the realistic nature of dark energy may not be the cosmological constant $\Lambda$, and proposes a great challenge for theorists: is actually dark energy the cosmological constant or time-dependent physical components?

In general, dark energy is parameterized by an equation of state (EoS) $\omega(z)=p_{de}/\rho_{de}$, where $z$, $\rho_{de}$ and $p_{de}$ denote the redshift, its energy density and pressure, respectively. In order to cope with the above-mentioned challenge, a great deal of efforts have been implemented through several cosmic probes, such as Type Ia supernovae (SNe Ia), baryon acoustic oscillations (BAO), observational Hubble parameter $H(z)$ data, strong and weak gravitational lensing data (S-WGL), gamma-ray burst data (GRB), and so on. As usual, in the literature, there are two main methods to explore whether the dark energy EoS is a constant: the first one is constraining the specific dark energy models, such as phantom models, quintessence models and modified theories of gravity; the second one is to investigate whether there are deviations from the $\Lambda$CDM model by utilizing the model-independent methods, such as local regression smoothing (LRS) \cite{4}, principal component analysis (PCA) \cite{5}, gaussian processes (GP) \cite{6}.

In the present study, we would like to use the GP to carry out reconstruction processes by using different cosmic data sets. The GP is a fully Bayesian approach for smoothing data, and can exhibit directly a reconstruction of a function from data without assuming a specific parameterization of the function. Consequently, one can determine any cosmological quantity from the cosmic data, and the key requirement of GP is only the covariance function which entirely depends on the cosmic data. For this reason, previously, the model-independent GP has been widely applied into studying the expansion dynamics of the universe \cite{6,7,8}, the distance duality relation \cite{9}, the cosmography \cite{10}, the test of the $\Lambda$CDM model \cite{11}, the determination of the interaction between dark energy and dark matter \cite{12}, dodging the matter degeneracy to determine the dynamics of dark energy \cite{13}, the slowing down of cosmic acceleration \cite{14}, dodging the cosmic curvature to probe the constancy of the speed of light \cite{15}, and so forth.

The rest of this study is outlined in the following manner. In Section 2, we would like to review briefly on the GP methodology. In Section 3, we exhibit our reconstruction results. In the final section, discussions and conclusions are presented.

\section{GP Methodology}
In a Friedmann-Robertson-Walker (FRW) universe, the luminosity distance $d_L(z)$ can be written as
\begin{equation}
d_L(z)=\frac{c(1+z)}{H_0\sqrt{|\Omega_{k0}|}}sinn\left(\sqrt{|\Omega_{k0}|}\int^{z}_{0}\frac{dz'}{E(z')}\right), \label{1}
\end{equation}
where the dimensionless Hubble parameter $E(z)=H(z)/H_0$, the present-day cosmic curvature $\Omega_{k0}=-Kc^2/(a_0H_0^2)$, and for $sinn(x)= sin(x), x, sinh(x)$, $K=1, 0, -1$ , which corresponds to a closed, flat and open universe, respectively. In succession, using the normalized comoving distance $D(z)=(H_0/c)(1+z)^{-1}d_L(z)$, the dark energy EoS can be expressed as
\begin{equation}
\omega(z)=\frac{2(1+z)(1+\Omega_{k0})D''-[(1+z)^2\Omega_{k0}D'^2-3(1+\Omega_{k0}D^2)+2(1+z)\Omega_{k0}DD']D'}{3D'\{(1+z)^2[\Omega_{k0}+(1+z)\Omega_{m0}]D'^2-(1+\Omega_{k0}D^2)\}}, \label{2}
\end{equation}
where the prime denotes the derivative with respect to the redshift $z$ and $\Omega_{m0}$ is the present-day value of the matter density ratio parameter. It is worth noting that in our situation, the quantities $D, D', D''$ of Eq. (\ref{2}) can be obtained from the reconstruction process, and the values of the parameters $\Omega_{k0}, \Omega_{m0}$ will affect the final reconstruction results of the dark energy EoS.

We use the online package GaPP (Gaussian Processes in Python) to carry out our reconstruction.
As noted in \cite{6}, the GP can directly reconstruct a function from observed data without assuming a specific model or choosing a parameterization for the underlying function.
In general, the GP is a generalization of a Gaussian distribution, which is the distribution of a random variable, and exhibits a distribution over functions.
At each reconstruction point $x$, the reconstructed function $f(x)$ is a Gaussian distribution with a mean value and Gaussian error. The keynote of the GP is a covariance function $k(x,\tilde{x})$ which correlates the function $f(x)$ at different reconstruction points. To be more precise, the covariance function $k(x,\tilde{x})$ depends only on two hyperparameters $l$ and $\sigma_f$, which characterize the coherent scale of the correlation in $x$-direction and typical change in the $y$-direction, respectively. Generally speaking, the choice is the squared exponential covariance function $k(x,\tilde{x})=\sigma_f^2 \mathrm{exp}[-|x-\tilde{x}|^2/(2l^2)]$. However, the analysis in \cite{16} has verified that the Mat\'{e}rn ($\nu=9/2$) covariance function is a better choice to implement the reconstruction. Thus, we would like to adopt the Mat\'{e}rn ($\nu=9/2$) covariance function in the following analysis:
\begin{equation}
k(x,\tilde{x})=\sigma_f^2 \mathrm{exp}(-\frac{3|x-\tilde{x}|}{l})\times[1+\frac{3|x-\tilde{x}|}{l}+\frac{27(x-\tilde{x})^2}{7l^2}+\frac{18|x-\tilde{x}|^3}{7l^3}+\frac{27(x-\tilde{x})^4}{35l^2}]. \label{3}
\end{equation}
This indefinitely differentiable function is very useful to reconstruct the derivative of a specific function.
In our reconstruction, we use the Union 2.1 data set \cite{17} which contains 580 SNe Ia data points. We also transform the theoretical distance modulus $m-M$ to $D$ in the following manner
\begin{equation}
m-M-25+5\lg(\frac{H_0}{c})=5\lg[(1+z)D]. \label{4}
\end{equation}
As described in the previous literature \cite{6}, we set the initial conditions $D(z=0)$ and $D'(z=0)=1$ in the reconstruction process. It is noteworthy that the values of $D$ depend only on a combination of the absolute magnitude $M$ and the present-day the Hubble parameter $H_0$.

Different from the previous literature \cite{6}, we have compiled the 30 latest $H(z)$ cosmic chronometer data points shown in Table. \ref{t1} and the CMB shift parameter $\mathcal{R}=1.7488\pm0.0074$ from the recent Planck's release \cite{18} as important supplements of SNe Ia data, since the observations of the first derivative of $D$ and the CMB high-redshift observation have been not used in the literature. For the purposes to see more apparently, we exhibit the main parts of the original code as follows\\

\begin{figure}
\centering
\includegraphics[scale=0.5]{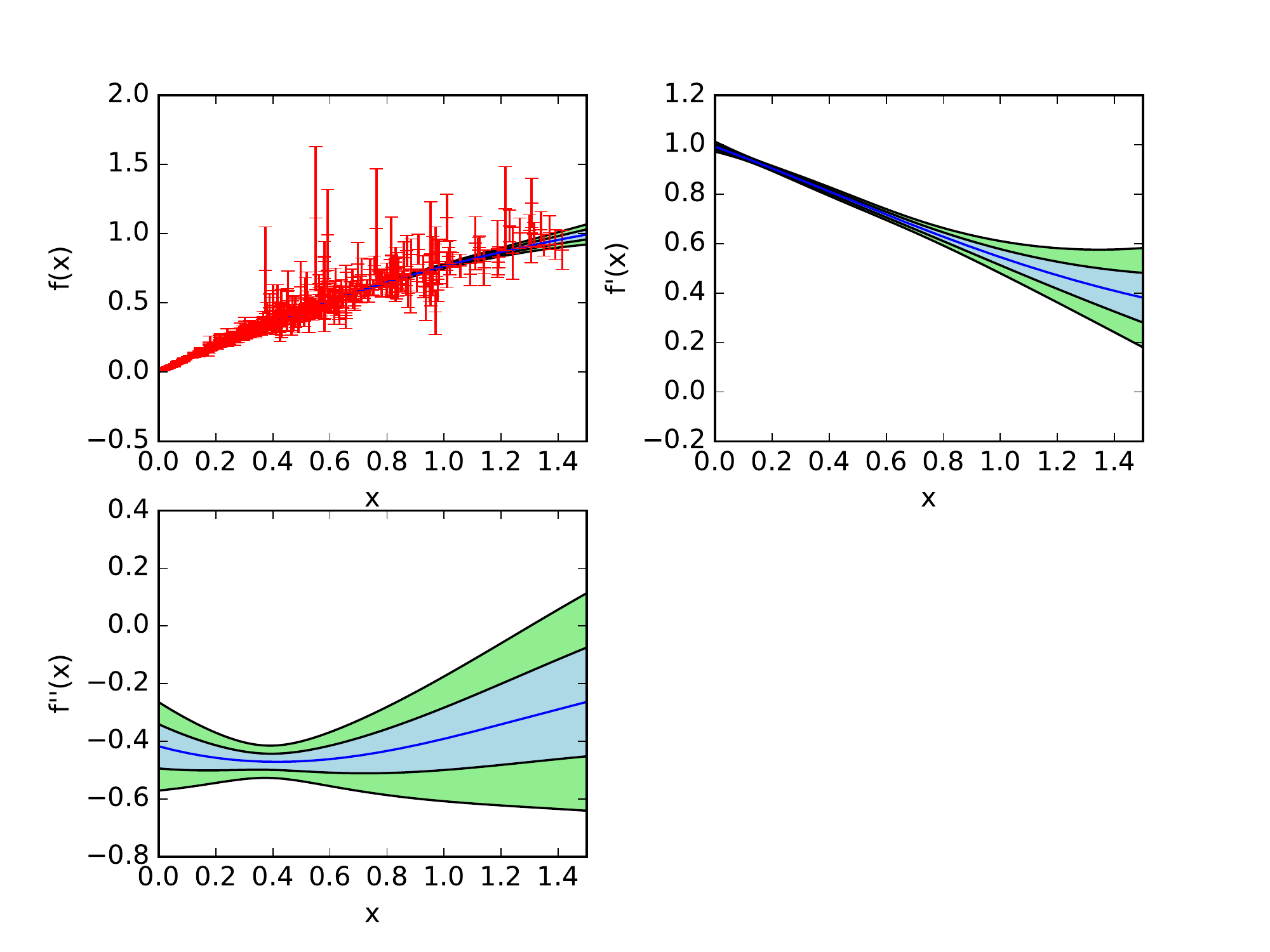}
\caption{The GP reconstructions of $f(x), f'(x)$ and $f''(x)$ by using SNe Ia. The shaded regions are reconstructions with $68\%$ and $95\%$ confidence level. The blue lines represents the underlying true model (the mean value of reconstructions). Since this figure is aimed at demonstrating the correctness of our GP reconstruction, we take the labels $x, f(x), f'(x)$ and $f''(x)$ as distinctions with those $z, D(z), D'(z)$ and $D''(z)$ used in Fig. \ref{f2}.}\label{f1}
\end{figure}

$\star$$dgp$.DGaussianProcess($X$, $Y$, $Sigma$, $covfunction=covariance.SquaredExponential$, $\fbox{dX}$, $\fbox{dY}$, $\fbox{dSigma}$),\\

where $X$, $Y$ and $Sigma$ denote the horizontal coordinate, longitudinal coordinate, statistical error of observational data of the reconstructed function, respectively; $dX$, $dY$ and $dSigma$ represent the horizontal coordinate, longitudinal coordinate, statistical error of observational data ($H(z)$ data) corresponding to the first derivative of the reconstructed function, respectively; the boxes correspond to the newly added data, which is considered in the literature about GP for the first time. Furthermore, the relations between the normalized comoving distance $D$ and the above-mentioned data can be expressed as
$$ relations\Longrightarrow\left\{
\begin{aligned}
D & \Longrightarrow   m-M \\
D & \Longrightarrow  \mathcal{R}=\sqrt{\Omega_{m0}}\int^{z_c}_0\frac{dz'}{E(z')} \\
D' & \Longrightarrow  \frac{H_0}{H(z)}
\end{aligned}
\right.
$$
where $z_c=1089.0$ is the redshift of recombination. Subsequently, we modify the code and use the Union 2.1 data set to test the correctness of our GP reconstruction. It is easy to be checked that our result in Fig. \ref{f1} is the same with that in Fig. 6 of \cite{14}.

\begin{table}[h!]
\begin{tabular}{ccccccc}
\hline
\hline
                          &$z$            & $H(z)$        & Ref.\\
\hline
                          &$0.070$        &$69\pm19.6$    & \cite{h1}                     \\
                          &$0.090$        &$69\pm12$      & \cite{h2}                    \\
                          &$0.120$        &$68.6\pm26.2$  & \cite{h1}                      \\
                          &$0.170$        &$83\pm8$       & \cite{h3}                     \\
                          &$0.179$        &$75\pm4$       & \cite{h4}                     \\
                          &$0.199$        &$75\pm5$       & \cite{h4}                     \\
                          &$0.200$        &$72.9\pm29.6$    & \cite{h1}                     \\
                          &$0.270$        &$77\pm14$      & \cite{h3}                     \\
                          &$0.280$        &$88.8\pm36.6$    & \cite{h1}                     \\
                          &$0.352$        &$83\pm14$    & \cite{h4}                     \\
                          &$0.3802$        &$83\pm13.5$    & \cite{h5}                     \\
                          &$0.400$        &$95\pm17$    & \cite{h3}                     \\
                          &$0.4004$        &$77\pm10.2$    & \cite{h5}                     \\
                          &$0.4247$        &$87.1\pm11.2$    & \cite{h5}                     \\
                          &$0.4497$        &$92.8\pm12.9$    & \cite{h5}                     \\
                          &$0.4783$        &$80.9\pm9$    & \cite{h5}                     \\
                          &$0.480$        &$97\pm62$    & \cite{h6}                     \\
                          &$0.593$        &$104\pm13$    & \cite{h4}                     \\
                          &$0.680$        &$92\pm8$    & \cite{h4}                         \\
                          &$0.781$        &$105\pm12$    & \cite{h4}                         \\
                          &$0.875$        &$125\pm17$    & \cite{h1}                         \\
                          &$0.880$        &$90\pm40$    & \cite{h6}                         \\
                          &$0.900$        &$117\pm23$    & \cite{h3}                         \\
                          &$1.037$        &$154\pm20$    & \cite{h4}                         \\
                          &$1.300$        &$168\pm17$    & \cite{h3}                         \\
                          &$1.363$        &$160\pm33.6$    & \cite{h7}                         \\
                          &$1.430$        &$177\pm18$    & \cite{h3}                         \\
                          &$1.530$        &$140\pm14$    & \cite{h3}                         \\
                          &$1.750$        &$202\pm40$    & \cite{h3}                         \\
                          &$1.965$        &$186.5\pm50.4$    & \cite{h7}                         \\

\hline
\hline
\end{tabular}
\caption{The latest $H(z)$ cosmic chronometers data points from different surveys.}
\label{t1}
\end{table}

\section{The results}
In this section, we will explore carefully the effects of the present-day matter density ratio parameter $\Omega_{m0}$, the Hubble constant $H_0$ and the cosmic curvature $\Omega_{k0}$ on the dark energy EoS, respectively, by using a combination of Union 2.1 data set, 30 H(z) data points and Planck's shift parameter. In addition, we also investigate the recent $H_0$ tension utilizing our GP reconstructions.

\begin{figure}
\centering
\includegraphics[scale=0.5]{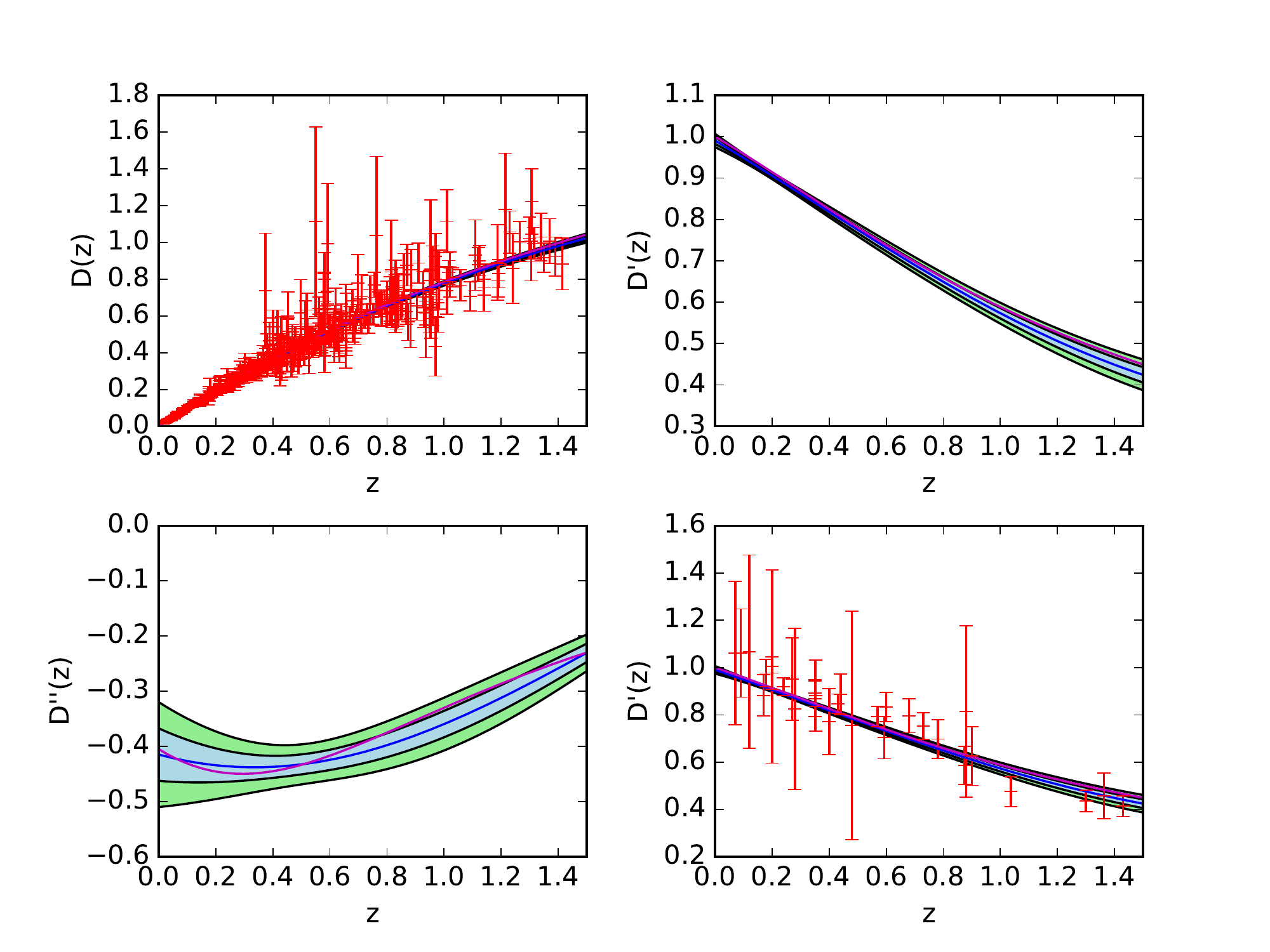}
\caption{The GP reconstructions of $D(z), D'(z)$ and $D''(z)$ using SNe Ia + H(z) + CMB. The blue and magenta lines represents the underlying true model (the mean value of reconstructions) and the $\Lambda$CDM model, respectively. The 30 $H(z)$ data points are exhibited in the lower right panel.}\label{f2}
\end{figure}

Using the recent Planck's result $\Omega_{m0}=0.308\pm0.012$ and the recent local value of the Hubble constant $73.24\pm1.74$ km s$^{-1}$ Mpc$^{-1}$ measured by Riess et al. \cite{19}, which is $3.4\sigma$ higher than the value of $66.93\pm0.62$ km s$^{-1}$ Mpc$^{-1}$ from CMB and BAO data analysis predicted by Planck collaboration \cite{18}, we find that the combined reconstructions $D(z), D'(z)$ and $D''(z)$ in Fig. \ref{f2} give out a tighter constraint than those in Fig. \ref{f1}, and that the base cosmological model is compatible with our reconstructions at $2\sigma$ level. To be more precise, the 30 newly added $H(z)$ data points give out a stricter constraint in the low-redshift range (see the lower right panel in Fig. \ref{f2}), and the Planck's shift parameter gives out a tighter high-redshift constraint avoiding the divergence in the high-redshift range of Fig. \ref{f1}. Furthermore, it is easy to find that our reconstructions of $D(z), D'(z)$ and $D''(z)$ give out a stricter constraint than those in Fig. 8 of \cite{6}, and is approximately consistent with the $\Lambda$CDM model at $1\sigma$ level.  Thus, we can reconstruct the dark energy EoS $\omega(z)$ better in terms of the stricter constraints on $D(z), D'(z)$ and $D''(z)$, and the corresponding result is shown in the upper left panel of Fig. \ref{f3}. One can easily find that the dark energy EoS is consistent with the base cosmological model at $2\sigma$ level.

\begin{figure}
\centering
\includegraphics[scale=0.4]{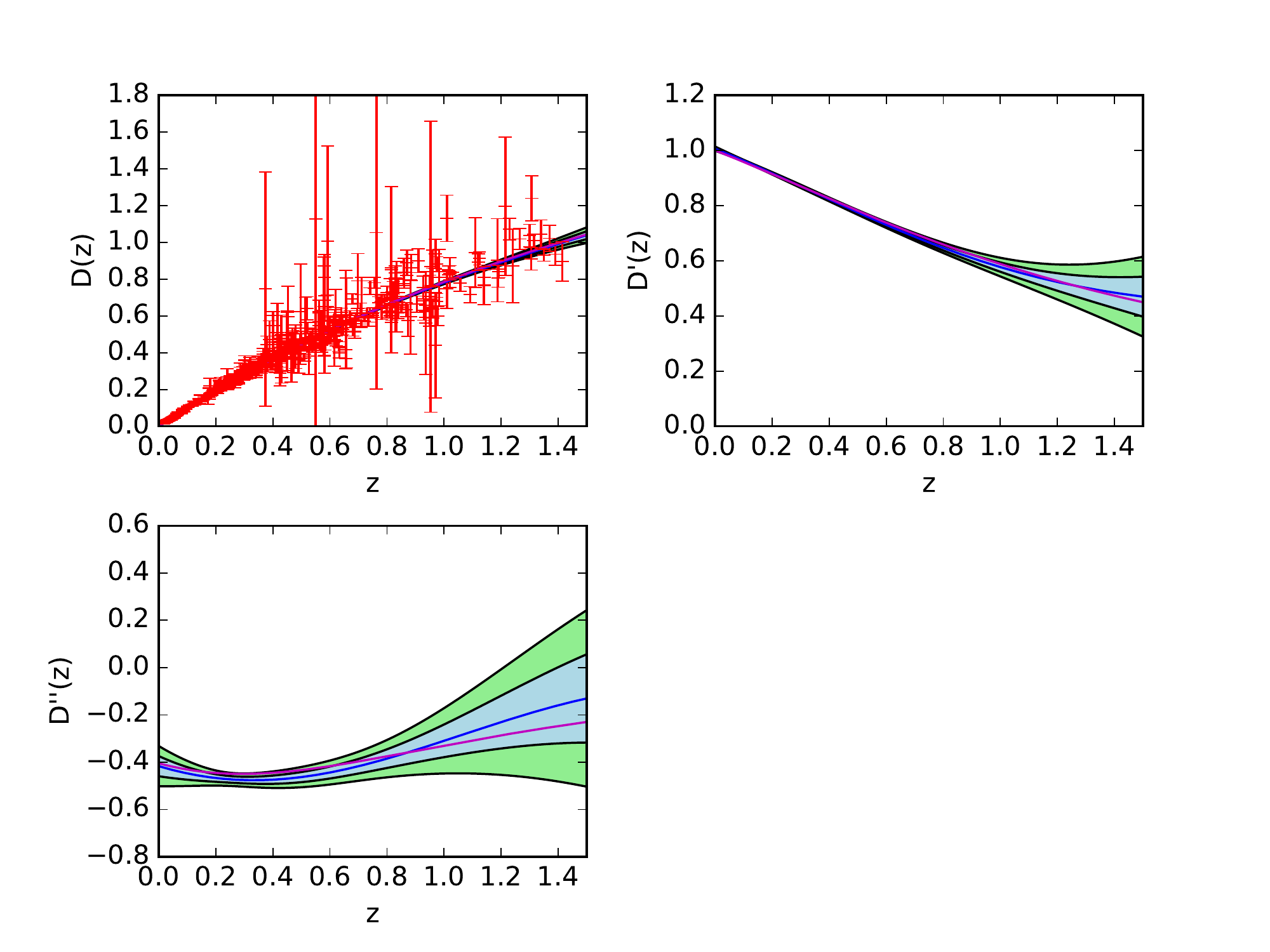}
\includegraphics[scale=0.4]{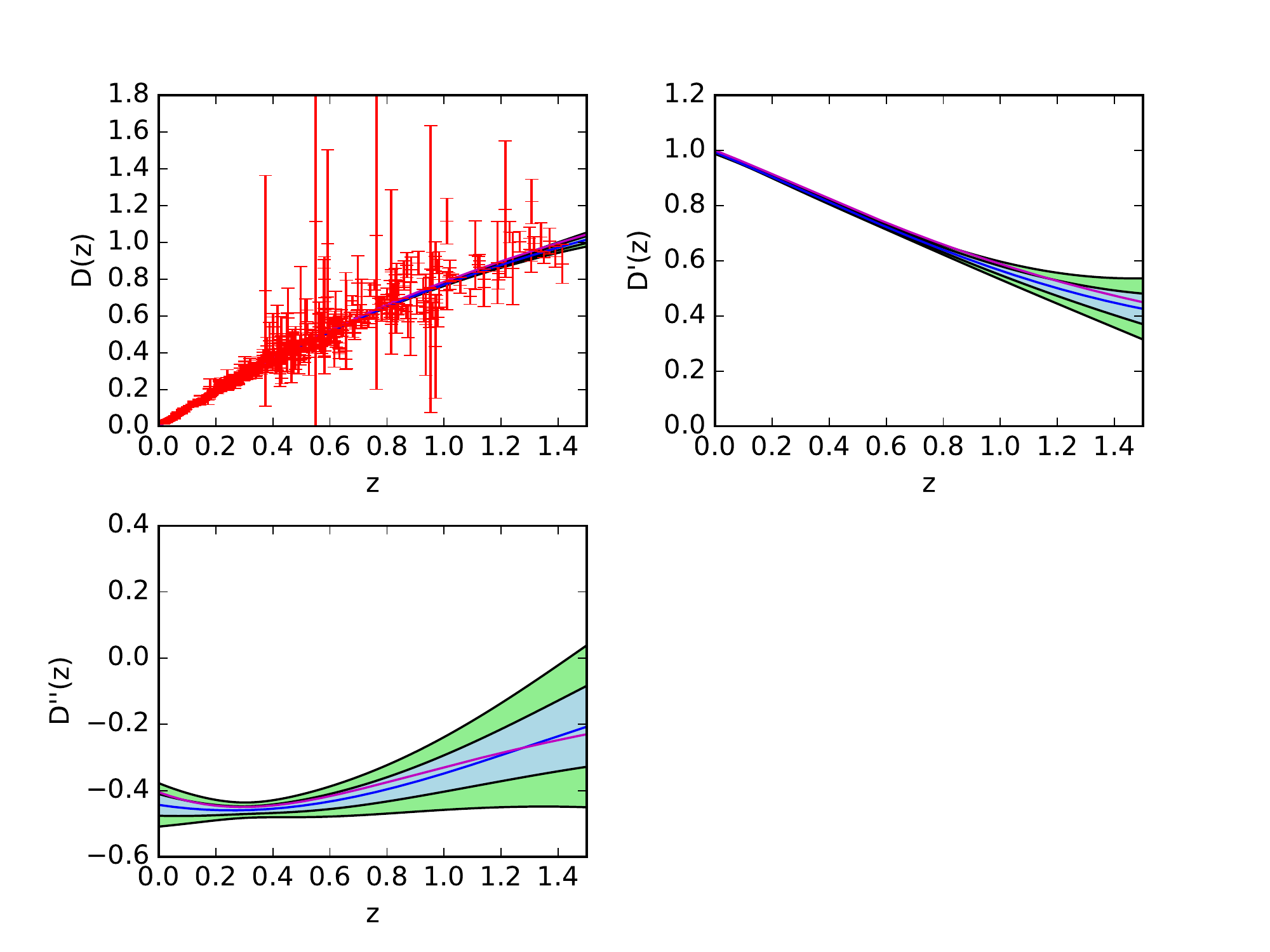}
\includegraphics[scale=0.4]{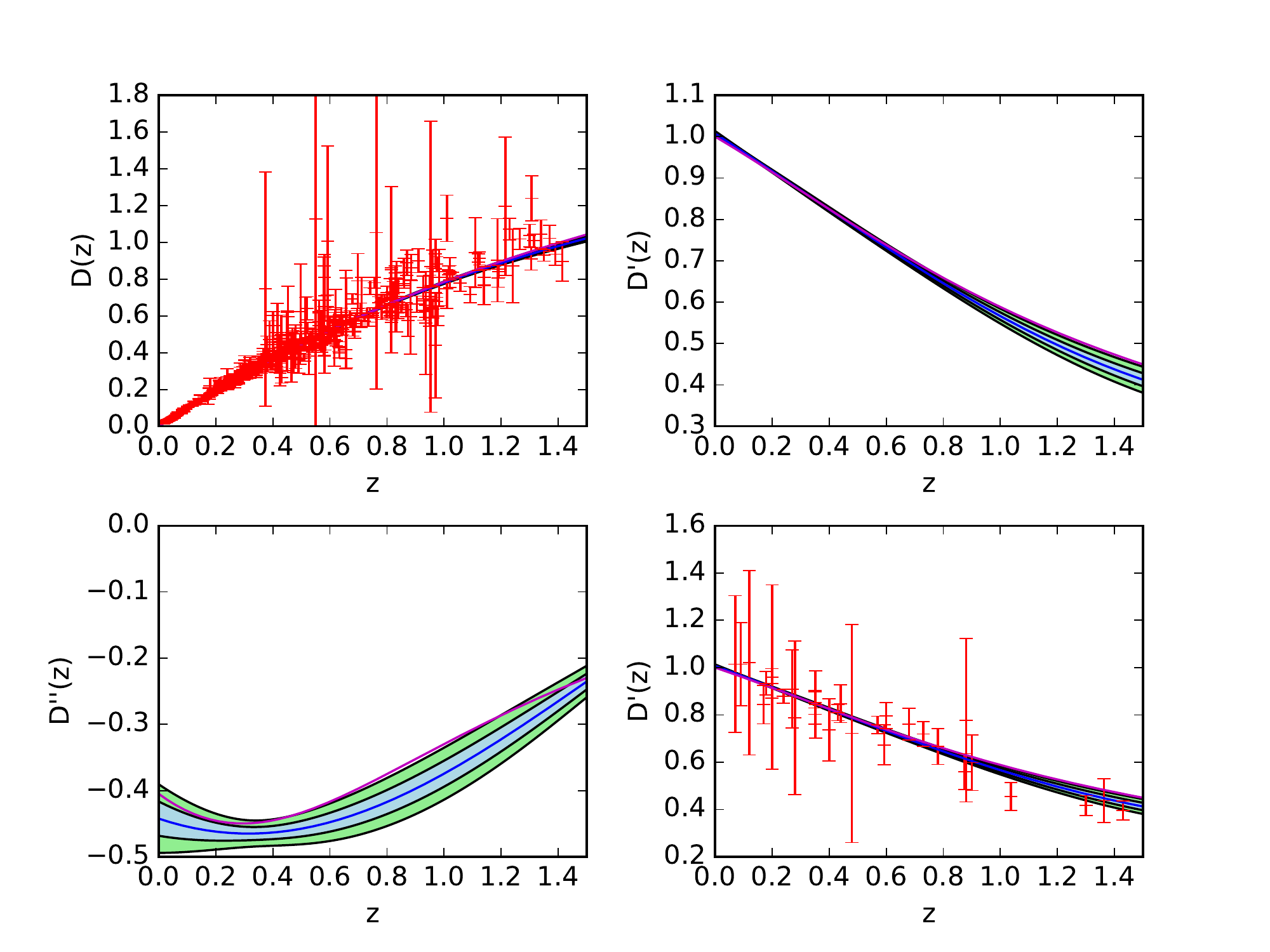}
\includegraphics[scale=0.4]{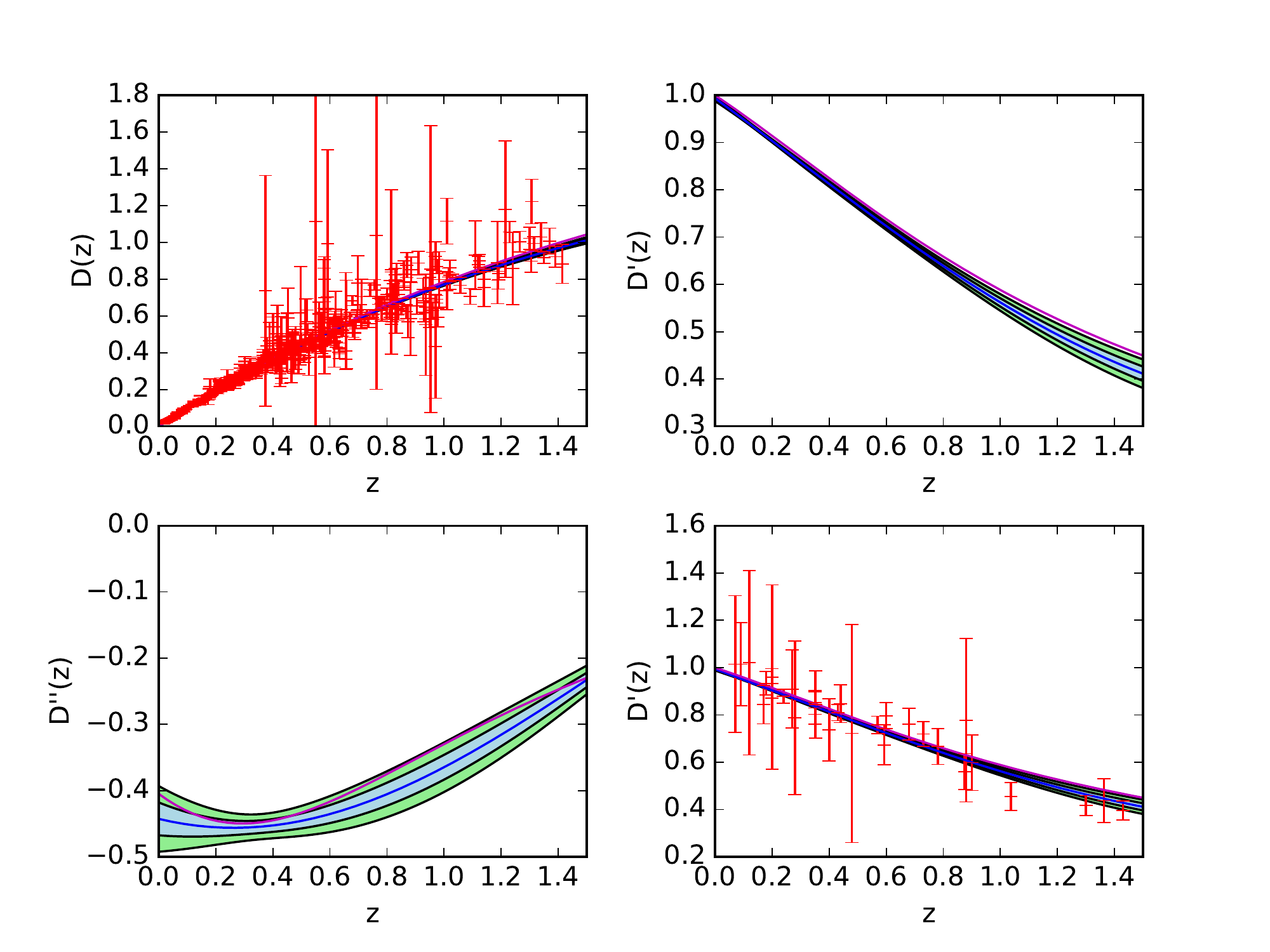}
\caption{The GP reconstructions of $D(z), D'(z)$ and $D''(z)$ using different observations. The upper left panel, upper right panel, lower left panel and lower right panel correspond to SNe Ia, SNe Ia + CMB, SNe Ia + H(z) and SNe Ia + H(z) + CMB, respectively. We have assumed $\Omega_{m0}=0.3$, $\Omega_{k0}=0$ and $H_0=70$ km s$^{-1}$ Mpc$^{-1}$.}\label{b1}
\end{figure}

\begin{figure}
\centering
\includegraphics[scale=0.2]{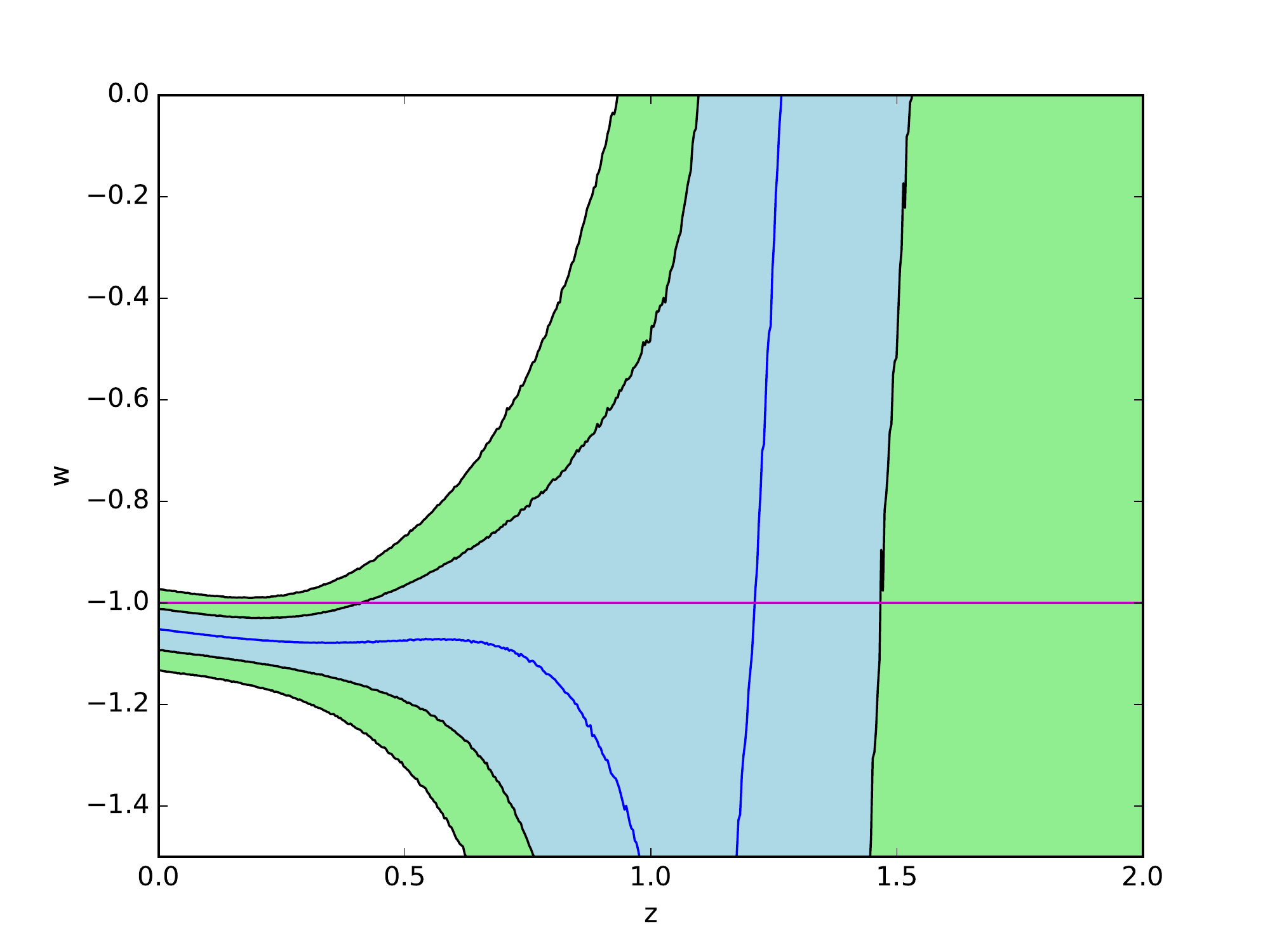}
\includegraphics[scale=0.2]{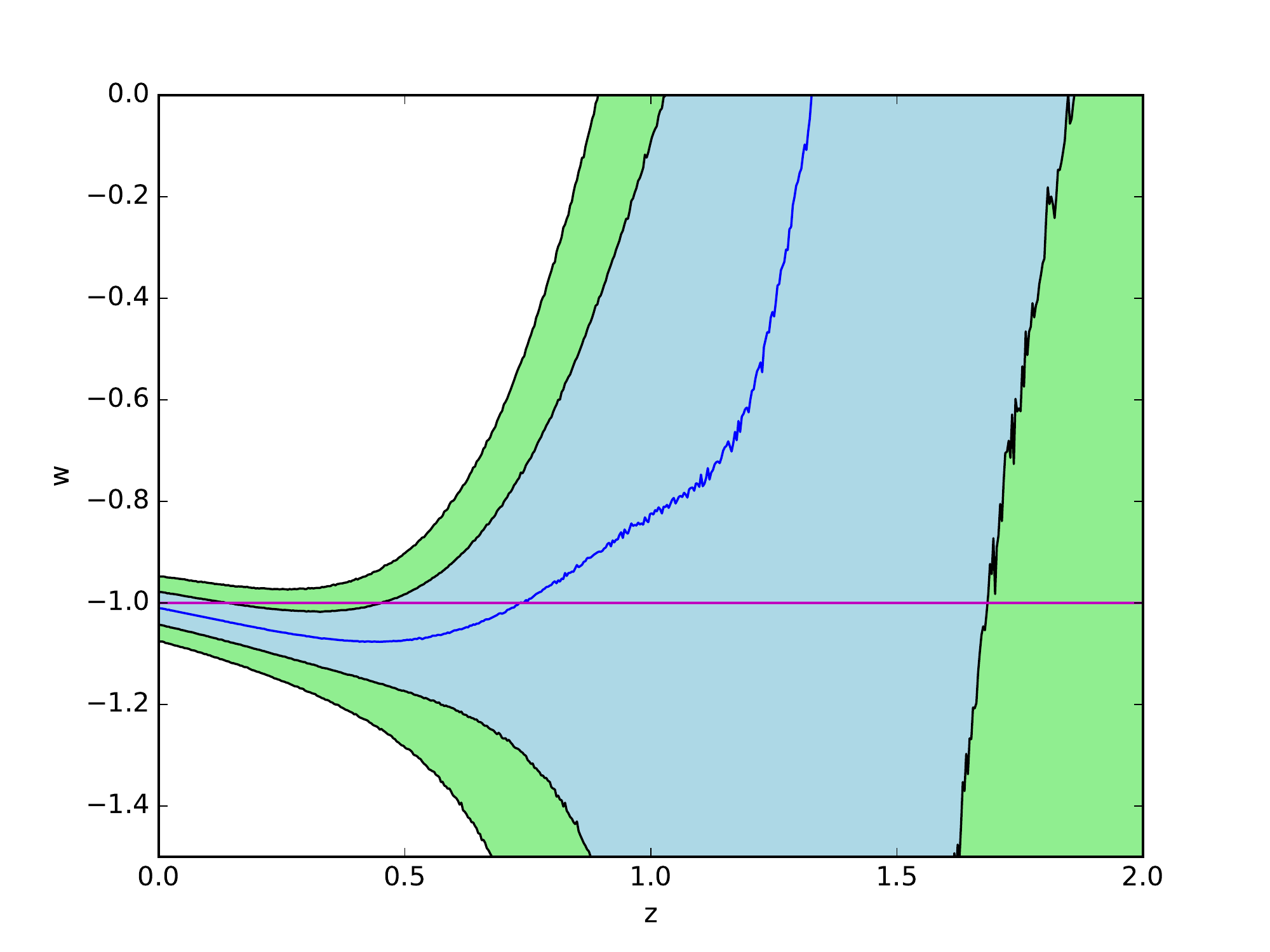}
\includegraphics[scale=0.2]{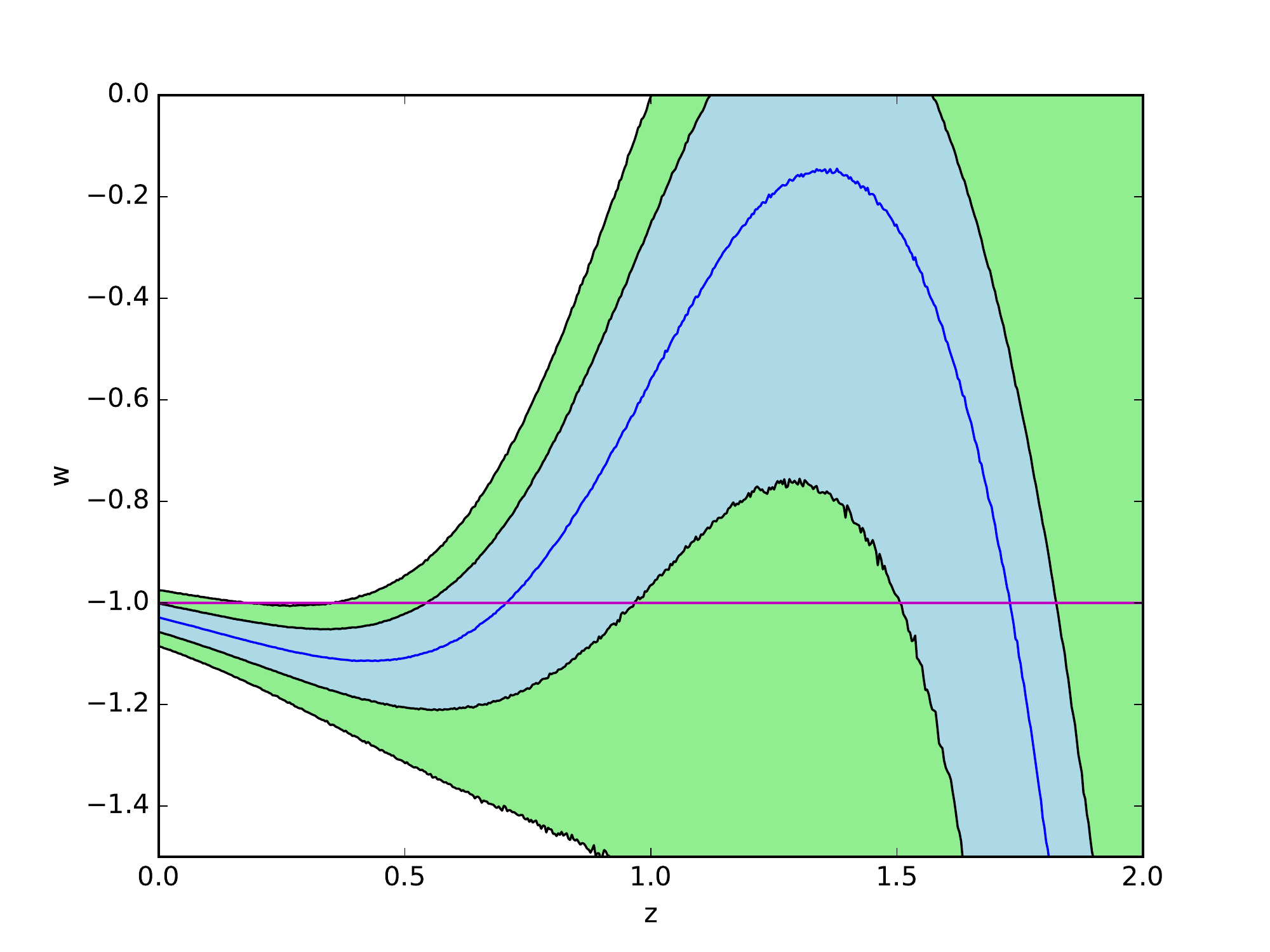}
\includegraphics[scale=0.2]{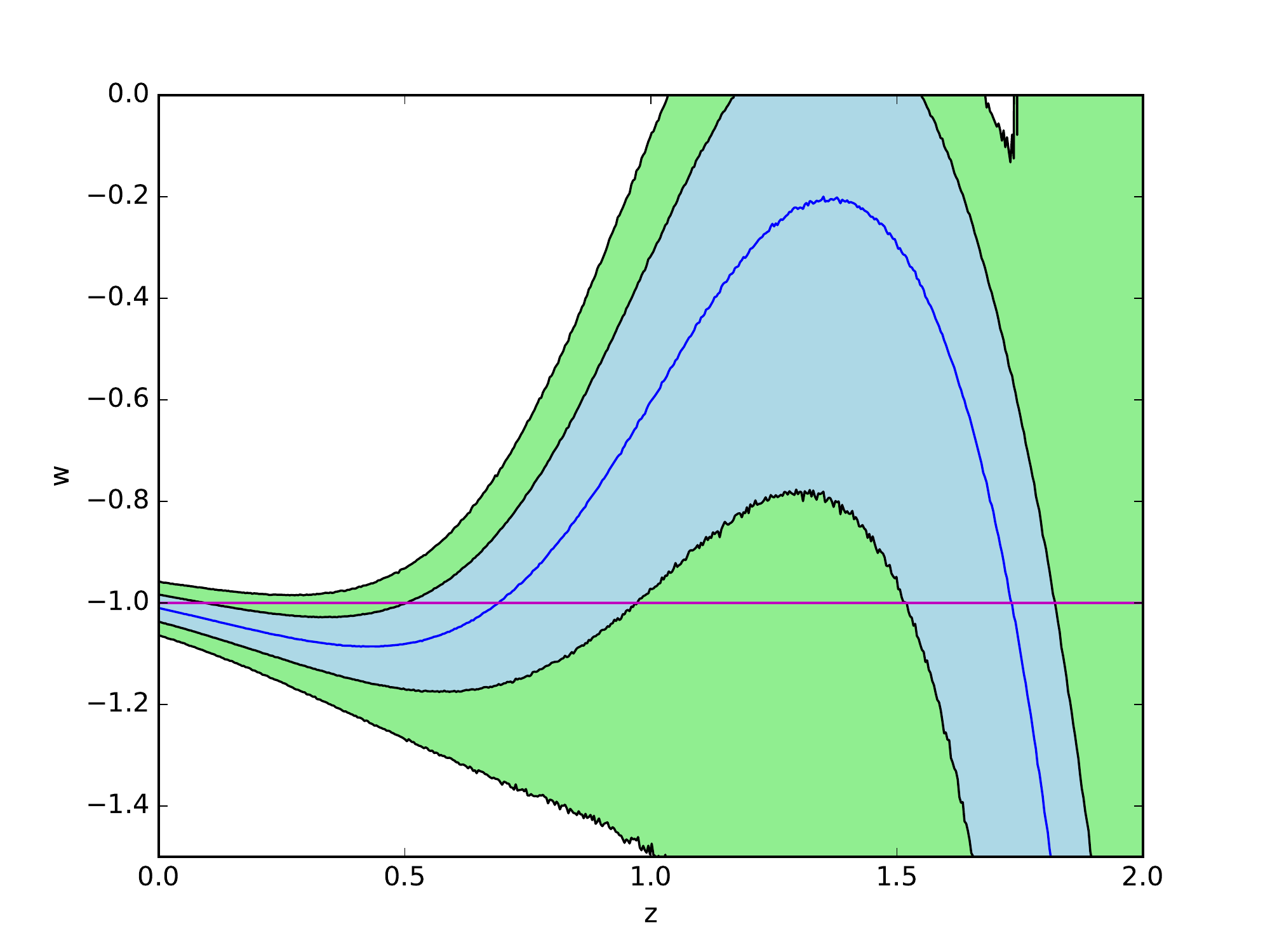}
\caption{The GP reconstructions of the dark energy EoS $\omega(z)$ using different observations. The different panels from left to right correspond to SNe Ia, SNe Ia + CMB, SNe Ia + H(z) and SNe Ia + H(z) + CMB, respectively. We have assumed $\Omega_{m0}=0.3$, $\Omega_{k0}=0$ and $H_0=70$ km s$^{-1}$ Mpc$^{-1}$.}\label{b2}
\end{figure}

\begin{figure}
\centering
\includegraphics[scale=0.23]{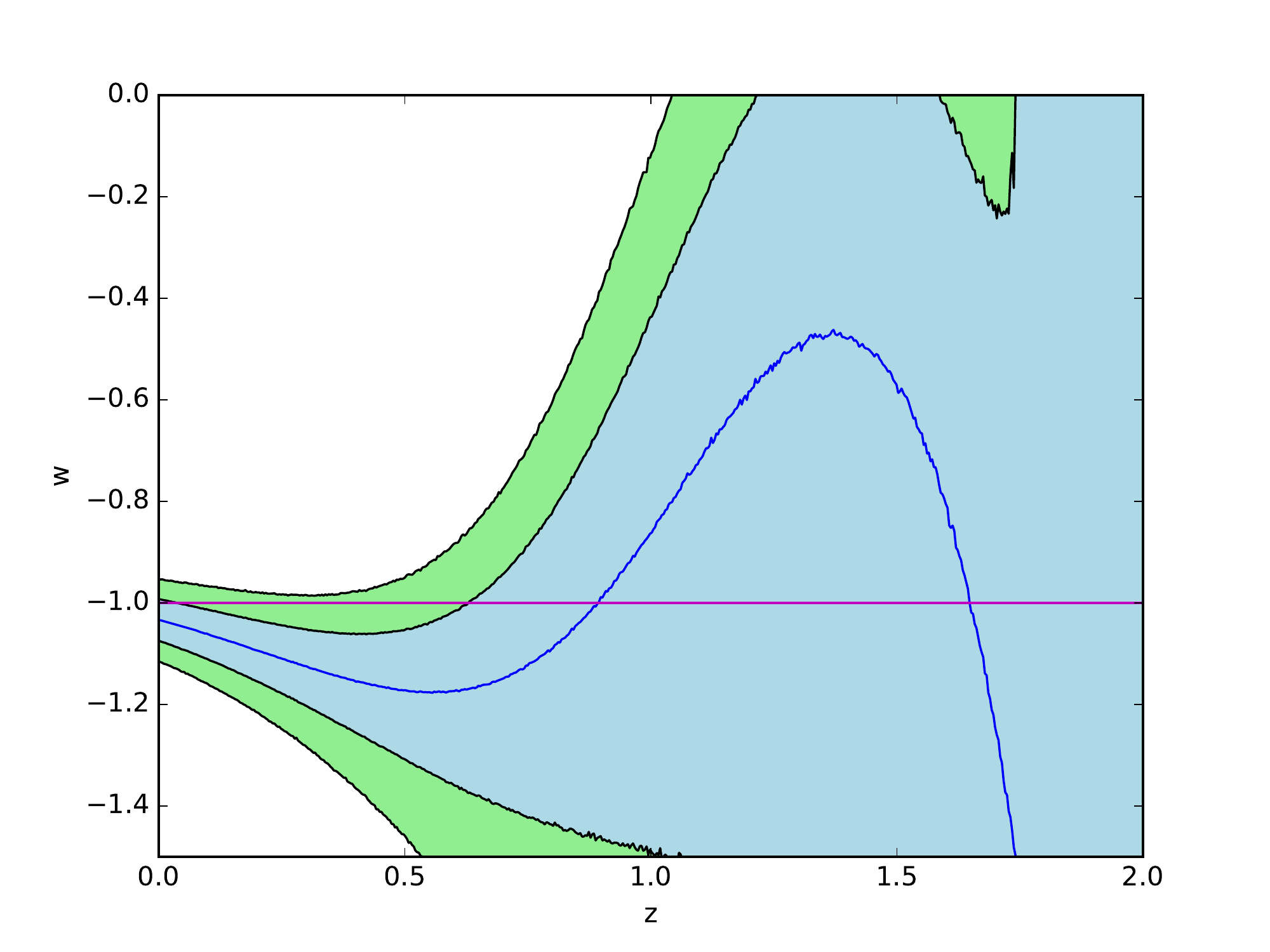}
\includegraphics[scale=0.23]{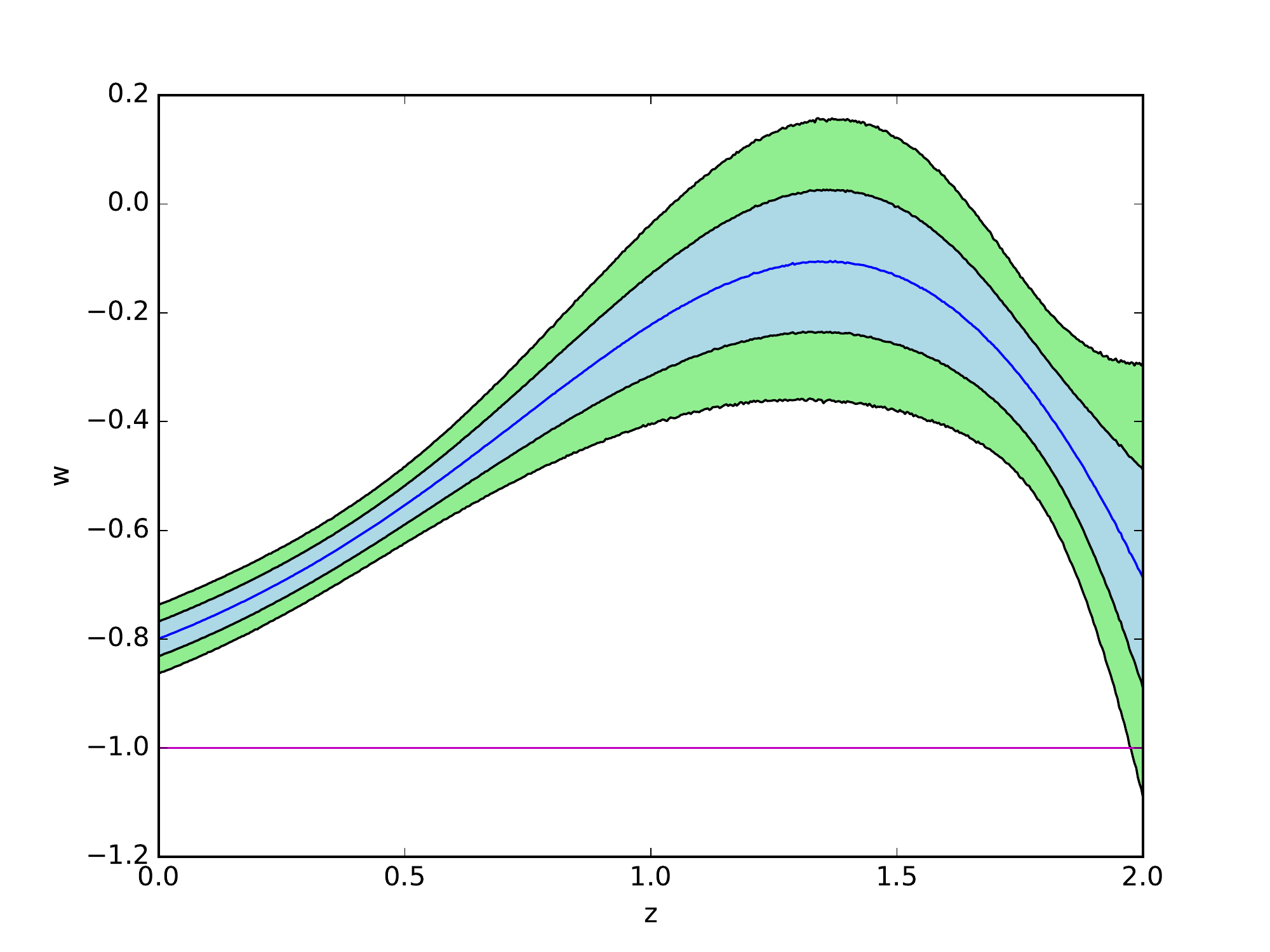}
\includegraphics[scale=0.23]{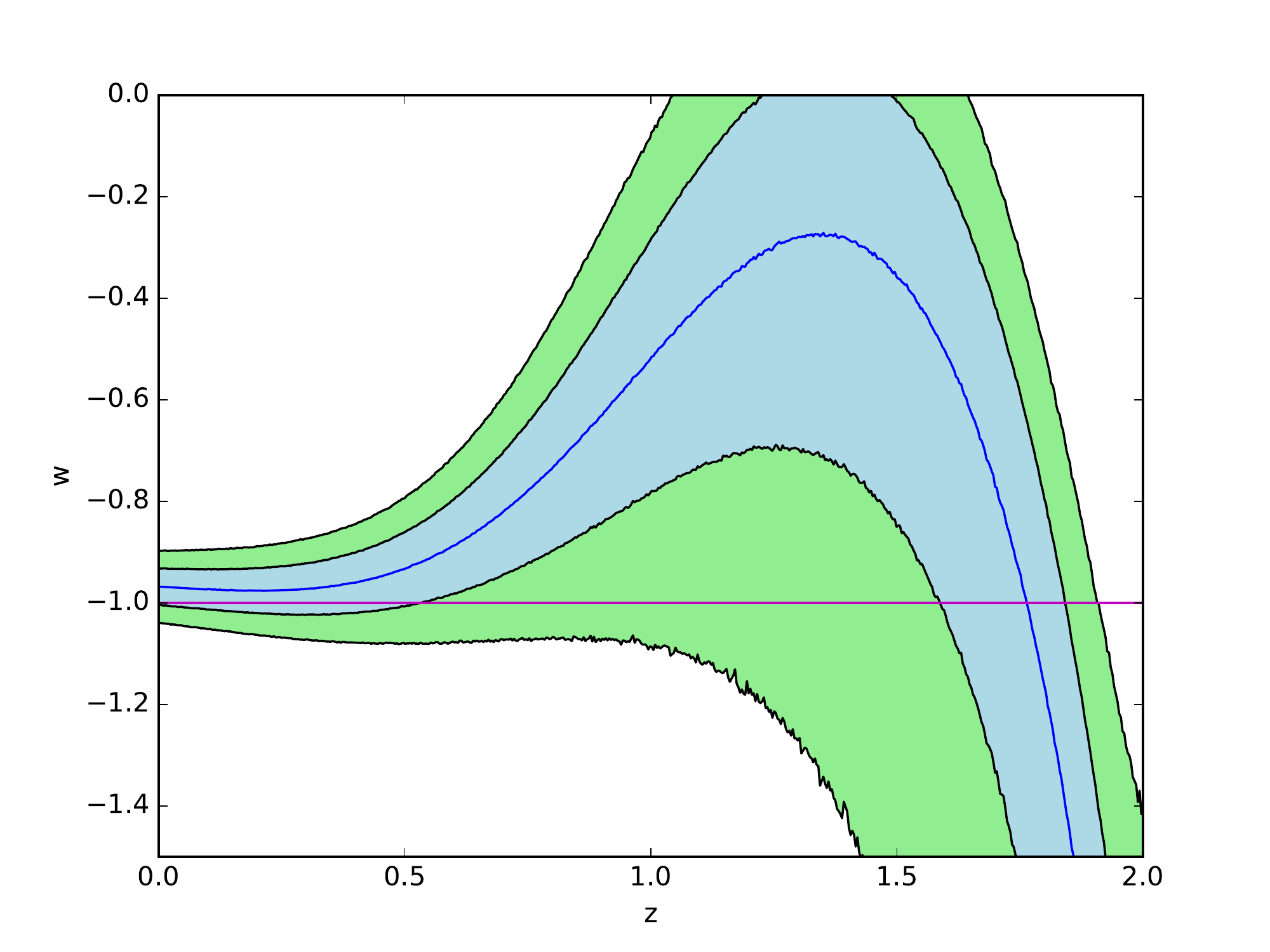}
\includegraphics[scale=0.23]{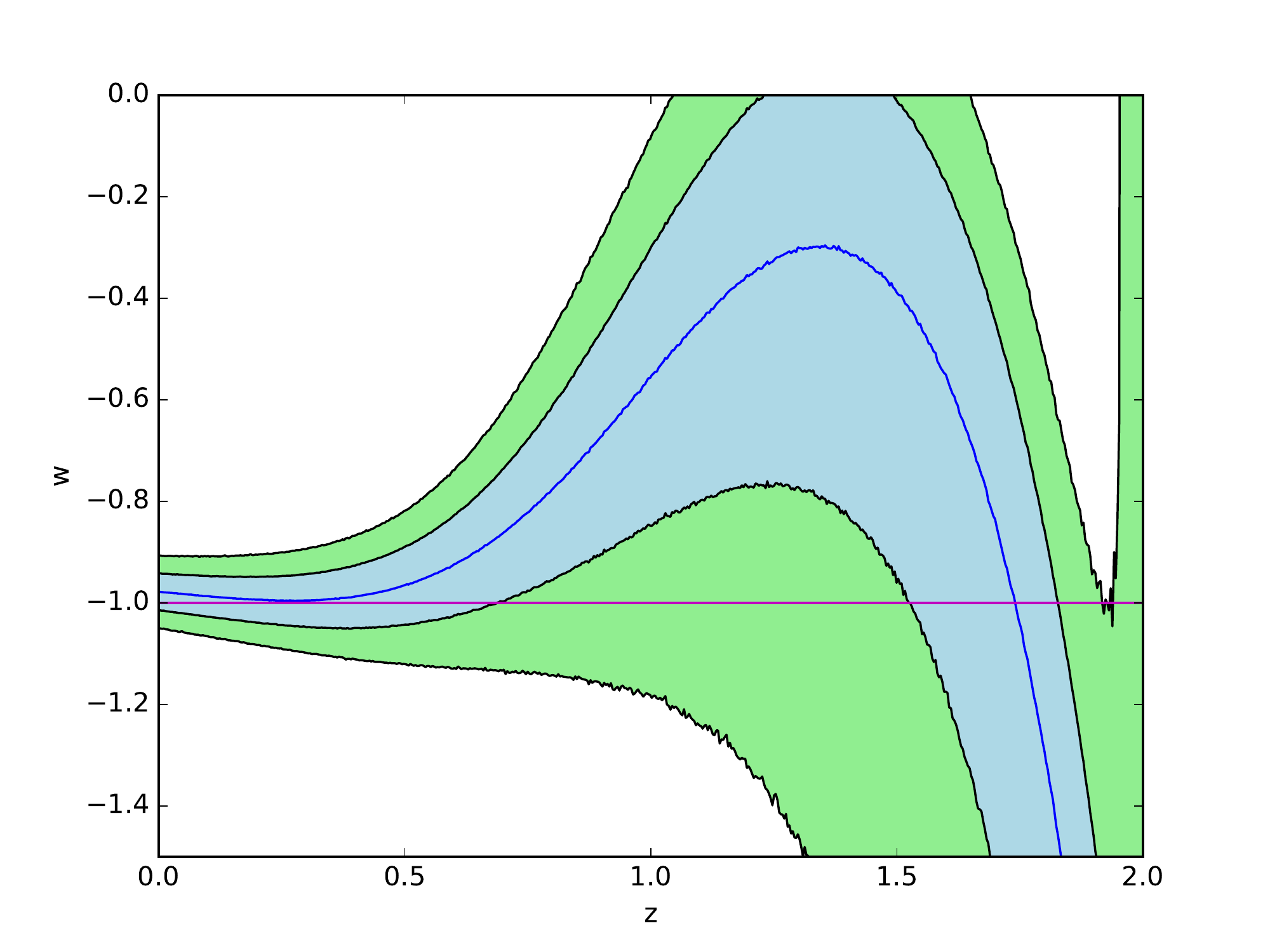}
\includegraphics[scale=0.23]{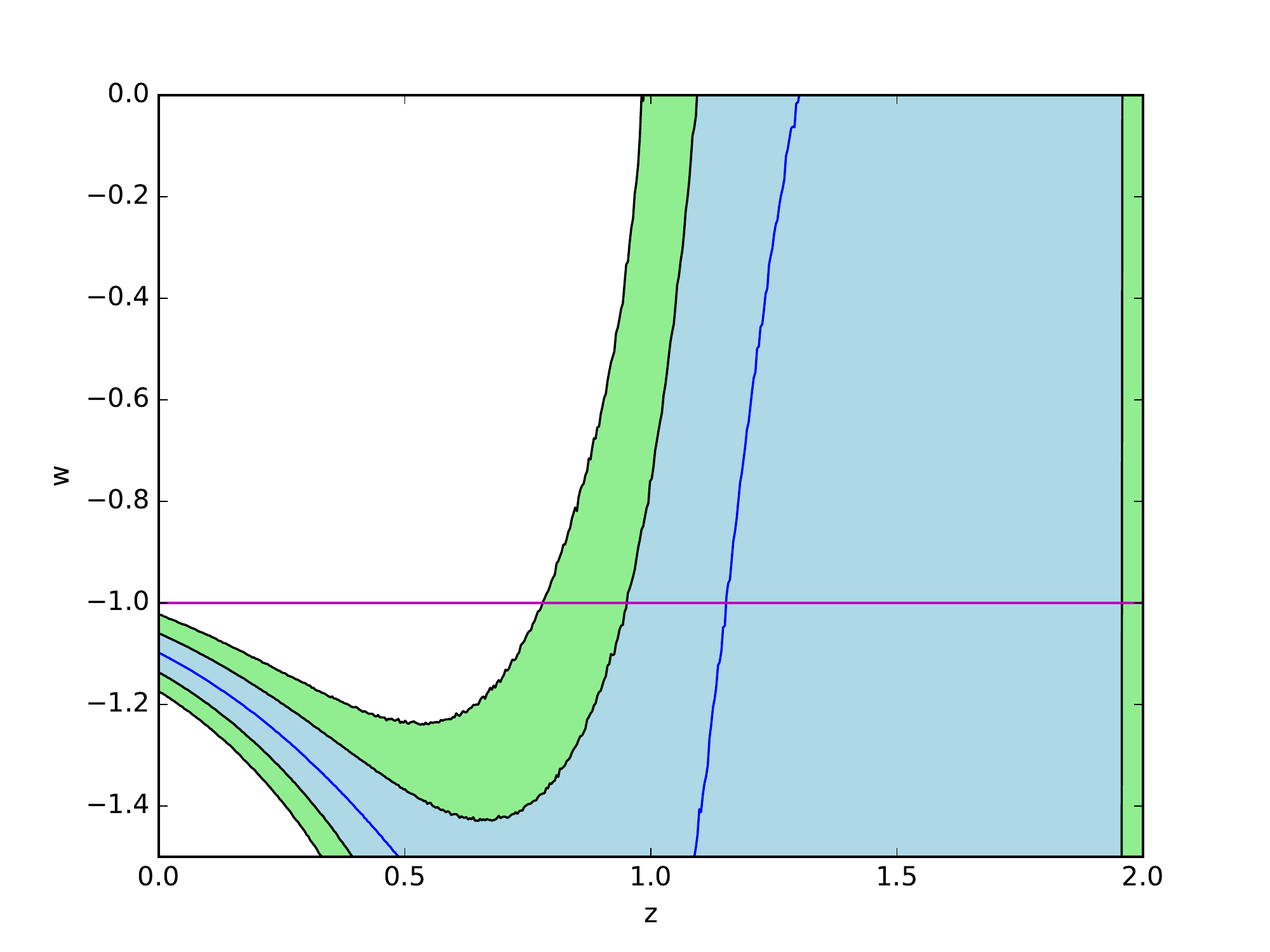}
\includegraphics[scale=0.23]{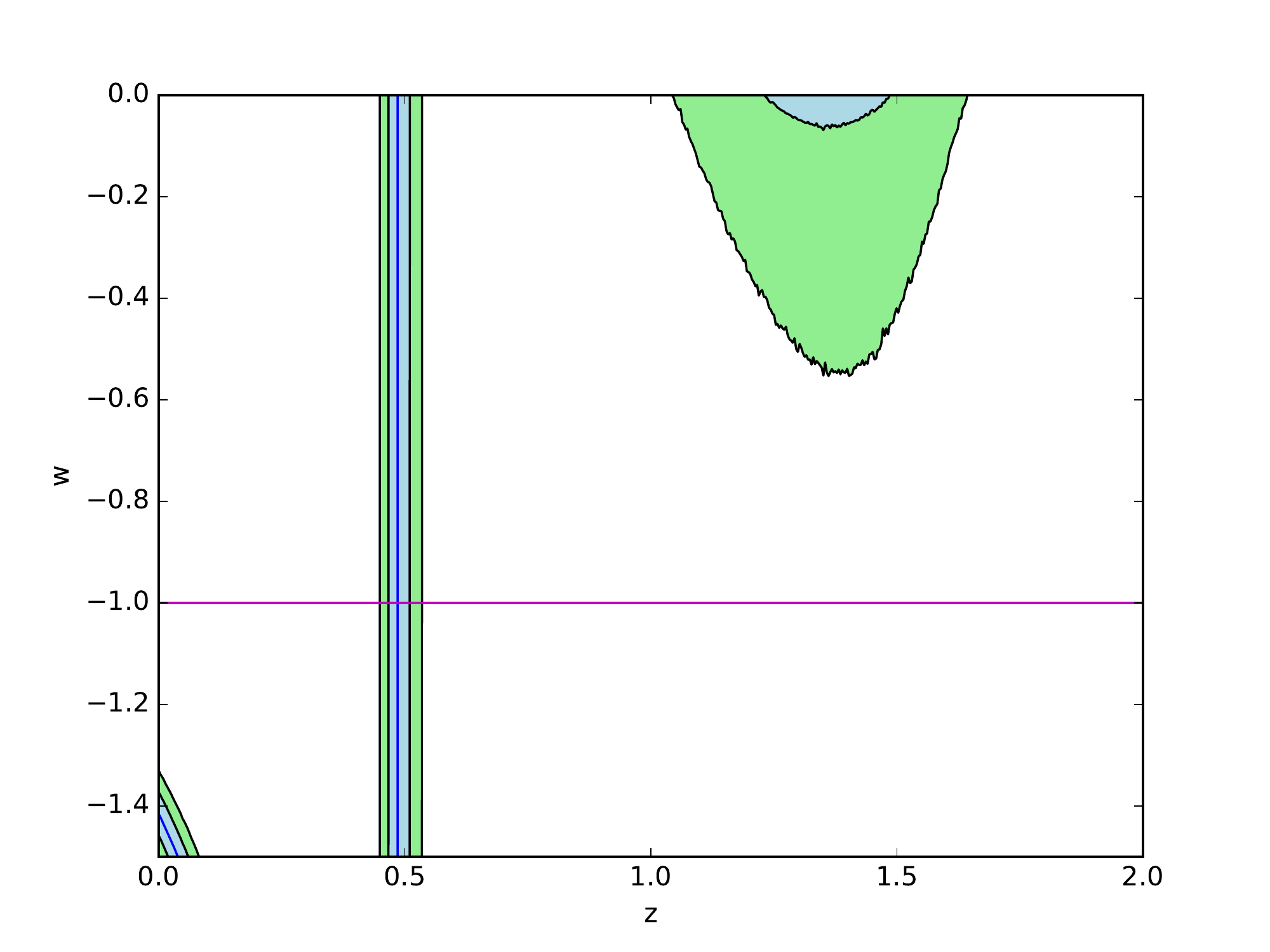}
\caption{The GP reconstructions of the dark energy EoS $\omega(z)$ using SNe Ia + H(z) + CMB. The upper panels from left to right correspond to the cases of $\Omega_{m0}=0.308\pm0.012$, $0.1$ and $0.26$, respectively. The lower panels from left to right correspond to the cases of $\Omega_{m0}=0.268$, $0.35$ and $0.5$, respectively. We have assumed $\Omega_{k0}=0$ and $H_0=73.24\pm1.74$ km s$^{-1}$ Mpc$^{-1}$.}\label{f3}
\end{figure}

To exhibit how much each probe is contributing better, we implement both the GP reconstructions of $D(z), D'(z)$ and $D''(z)$, and those of the dark energy EoS $\omega(z)$ using the above 3 different observations. From both the upper right panel of Fig. \ref{b1} and the second panel of Fig. \ref{b2}, one can find that the newly added CMB data just affects slightly the reconstructions of $D(z), D'(z)$, $D''(z)$ and the dark energy EoS in the low-redshift range, since it works well in the high-redshift range. However, from both the lower left panel of Fig. \ref{b1} and the third panel of Fig. \ref{b2}, one can find that the newly added H(z) probe improves apparently the reconstructions of $D(z), D'(z)$, $D''(z)$ and the dark energy EoS at $1\sigma$ level in the low-redshift range, and plays the main role in improved constraints on the dark energy EoS by using GP method. Note that here we have assumed the cosmic curvature $\Omega_{k0}=0$, and we would like to investigate the effects of $\Omega_{k0}$ on the dark energy EoS in the following subsection.

\subsection{The effects of variable $\Omega_{m0}$ on the dark energy EoS}
Notice that in Eq. (\ref{2}), the dark energy EoS depends obviously on two cosmological parameters $\Omega_{m0}$ and $\Omega_{k0}$ other than the data inputs $D(z), D'(z)$ and $D''(z)$ from GP, hence, we will study the effects of these two quantities on the dark energy EoS. Unlike the method proposed in \cite{13}, the authors construct a function $F(z)$ as a new null test to check the correctness of the flat $\Lambda$CDM model, in order to dodge the dark matter degeneracy while determining the dynamics of dark energy. In this study, we use the `` controlling variable method '' (change one parameter per time by leaving
others fixed) to investigate directly the effects of different parameters on the dark energy EoS. First of all, in this subsection, we consider the case of variable matter density ratio parameter by assuming $\Omega_{k0}=0$ and $H_0=73.24\pm1.74$ km s$^{-1}$ Mpc$^{-1}$.

One can easily find that, in the upper medium panel of Fig. \ref{f3}, the $\Lambda$CDM model lies out the $2\sigma$ confidence region apparently when $\Omega_{m0}=0.1$. This indicates that too small matter density ratio parameter can be directly ruled out using current cosmological observations. In the upper right panel of Fig. \ref{f3}, the $\Lambda$CDM model lies in the $2\sigma$ confidence region apparently when $\Omega_{m0}=0.26$, and the reconstruction appears to give out a tighter constraint in the low-redshift range than our standard case, i.e., $\Omega_{m0}=0.308\pm0.012$, $H_0=73.24\pm1.74$ km s$^{-1}$ Mpc$^{-1}$ and $\Omega_{k0}=0$ (see the upper left panel of Fig. \ref{f3}). However, this case deviates apparently from the $\Lambda$CDM model at $2\sigma$ confidence level when $z>1.9$. Subsequently, in the lower left panel of Fig. \ref{f3}, when $\Omega_{m0}=0.268$, one can find that the $\Lambda$CDM model happens to lie on the $2\sigma$ confidence region. In the two lower right panels of Fig. \ref{f3}, when $\Omega_{m0}=0.35$ and $0.5$, the reconstructions both deviates from the $\Lambda$CDM model at $2\sigma$ confidence level in the low-redshift range, which implies that too large matter density ratio parameter can be directly ruled out using currently cosmological observations. Interestingly, we find that our GP reconstruction can provide a relatively tight constraint on $\Omega_{m0}$ at $2\sigma$ confidence level, i.e., $\Omega_{m0}\in[0.268,0.310]$. To be more concrete, the $\Lambda$CDM model lies out the $2\sigma$ confidence region in the low-redshift range when $\Omega_{m0}>0.310$ (see the case $\Omega_{m0}=0.35$ in the lower medium panel of Fig. \ref{f3}), and in the relatively high-redshift range when $\Omega_{m0}<0.268$ (see the case $\Omega_{m0}=0.26$ in the upper right panel of Fig. \ref{f3}). In the meanwhile, the effects of variable $\Omega_{m0}$ on the reconstructions of $D(z), D'(z)$ and $D''(z)$ is too small to take into account them. From the point of view of observational data, $\Omega_{m0}$ is only associated with the Planck's shift parameter and the reconstructions of $D(z), D'(z)$ and $D''(z)$ is mainly determined by SNe Ia and $H(z)$ data. Consequently, its effects on the normalized comoving distance can be ignored.

Note that in the previous literature, the authors can not give out a relatively tight constraint on $\Omega_{m0}$ since they do not use the $H(z)$ and CMB data.

\subsection{The effects of variable $\Omega_{k0}$ on the dark energy EoS}
In this subsection, we continue using the `` controlling variable method '' to investigate directly the effects of variable cosmic curvature on the dark energy EoS by assuming $\Omega_{m0}=0.308\pm0.012$ and $H_0=73.24\pm1.74$ km s$^{-1}$ Mpc$^{-1}$.

\begin{figure}
\centering
\includegraphics[scale=0.2]{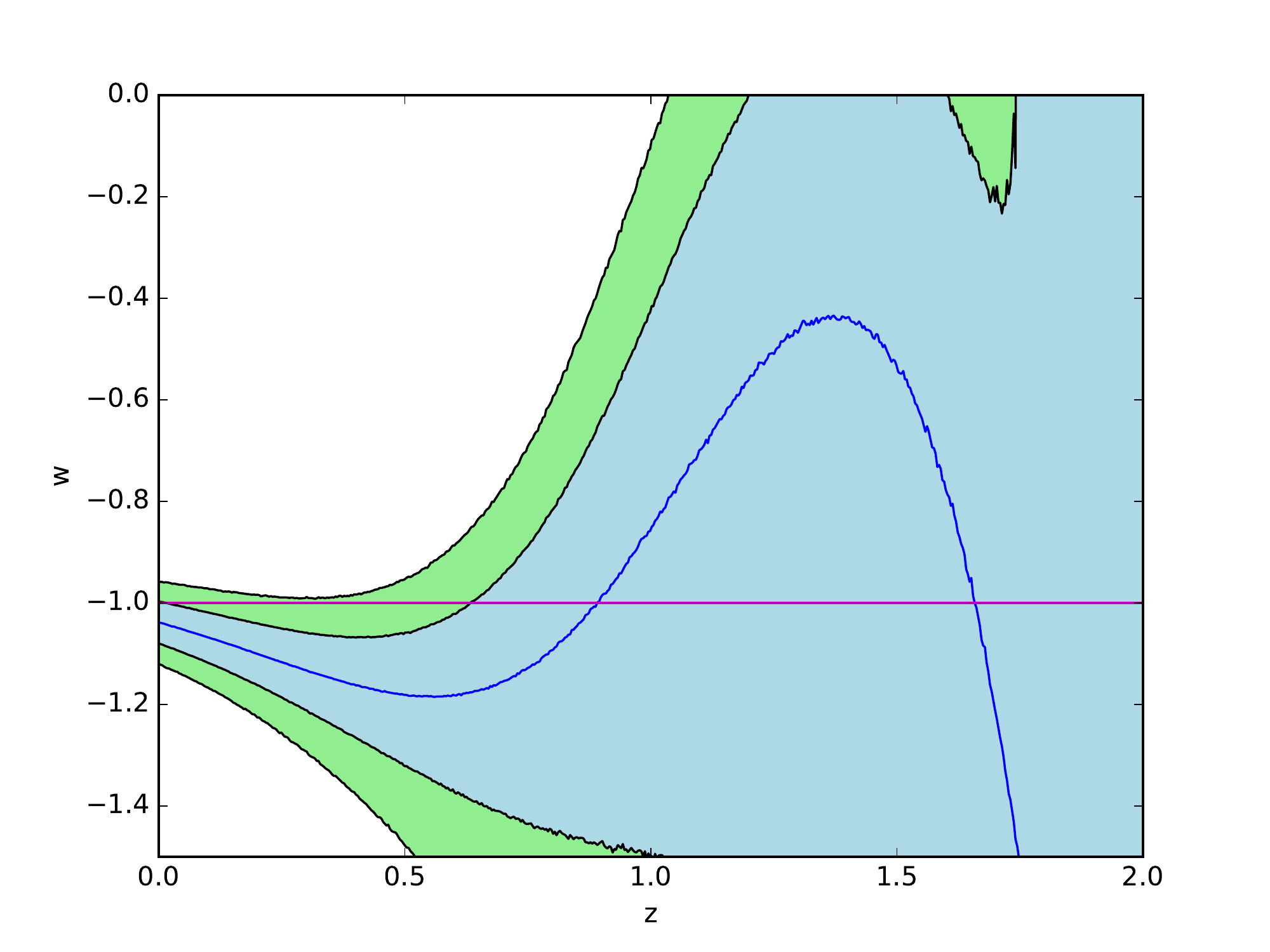}
\includegraphics[scale=0.2]{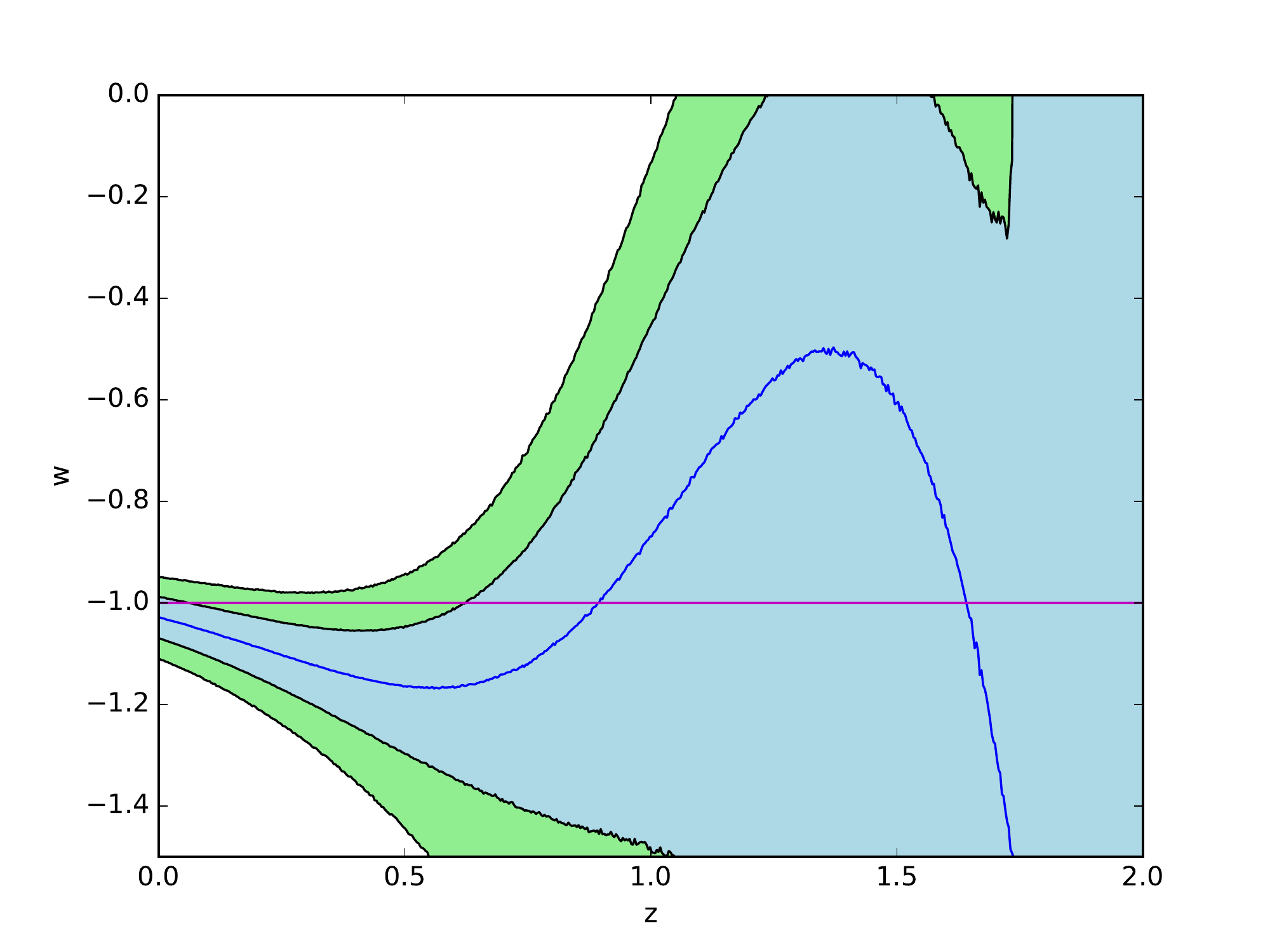}
\includegraphics[scale=0.2]{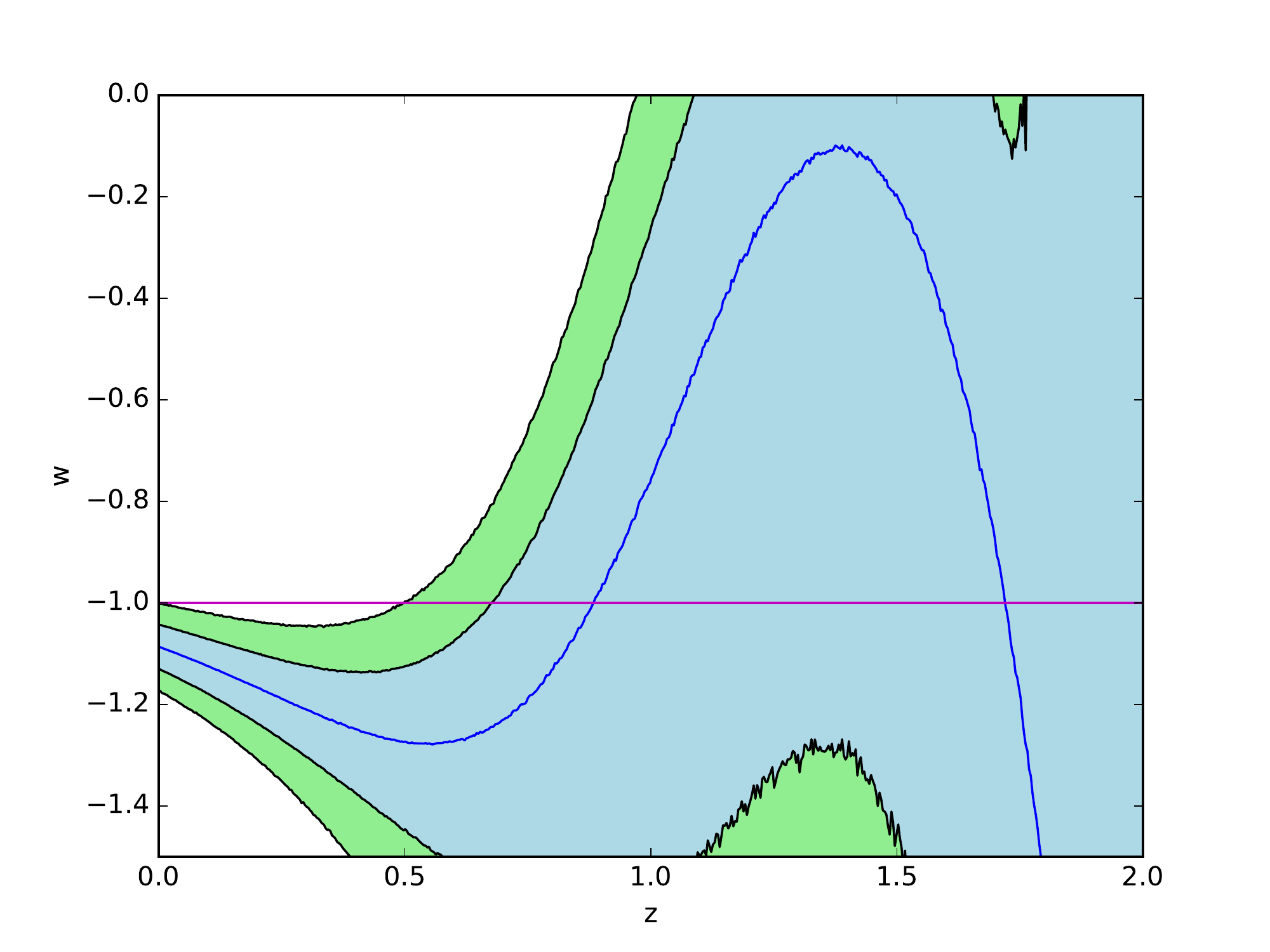}
\includegraphics[scale=0.2]{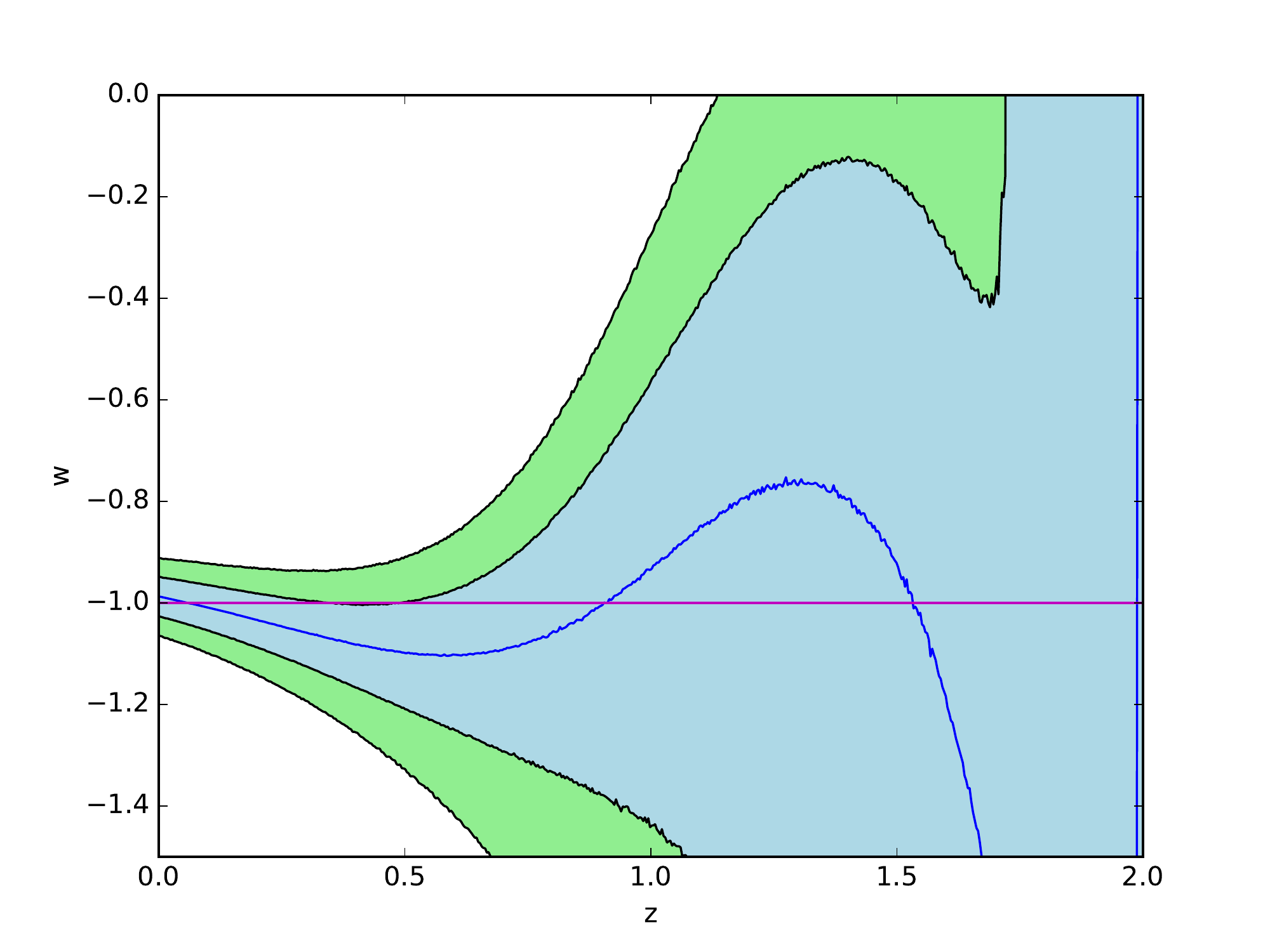}
\includegraphics[scale=0.2]{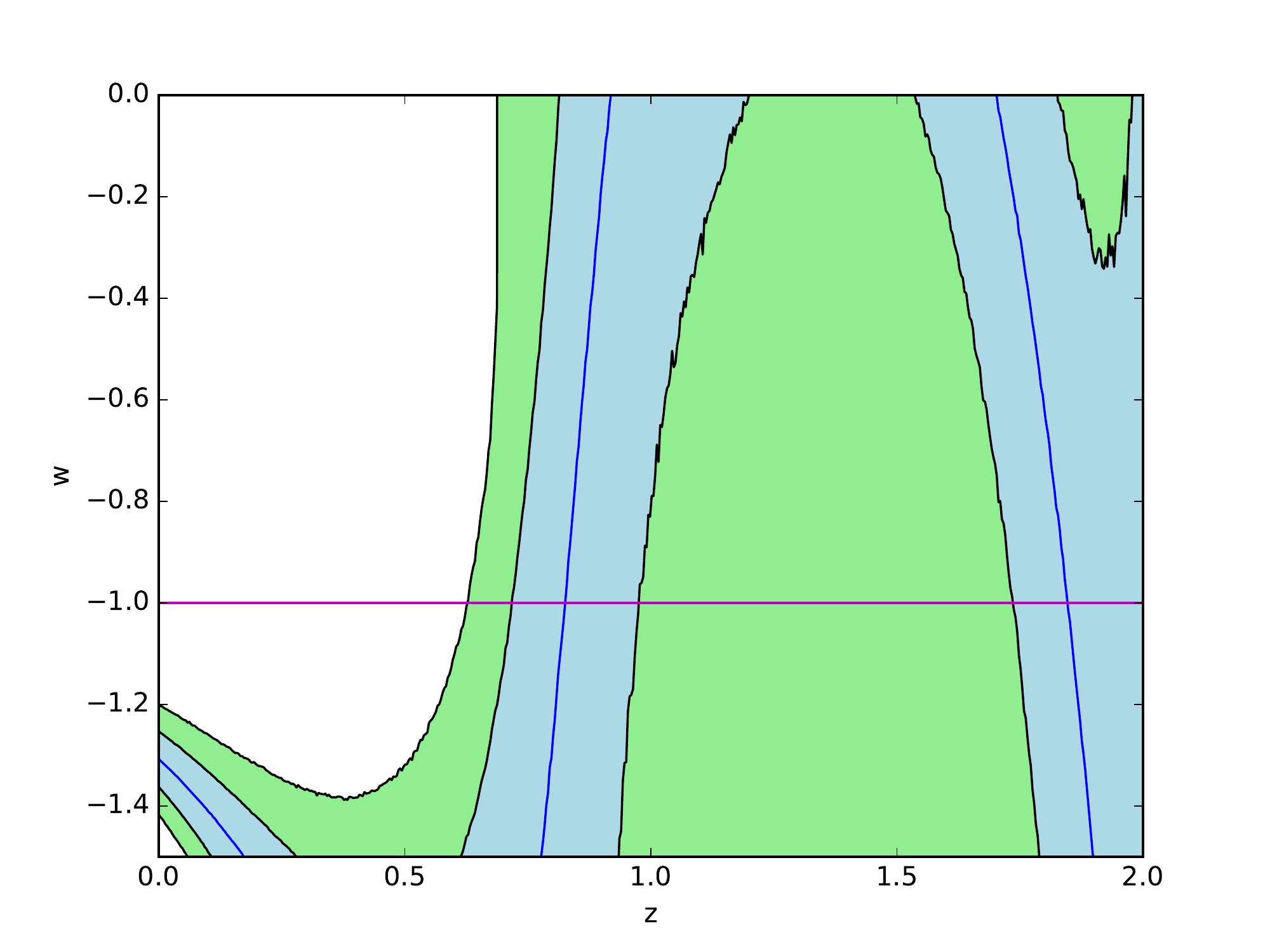}
\includegraphics[scale=0.2]{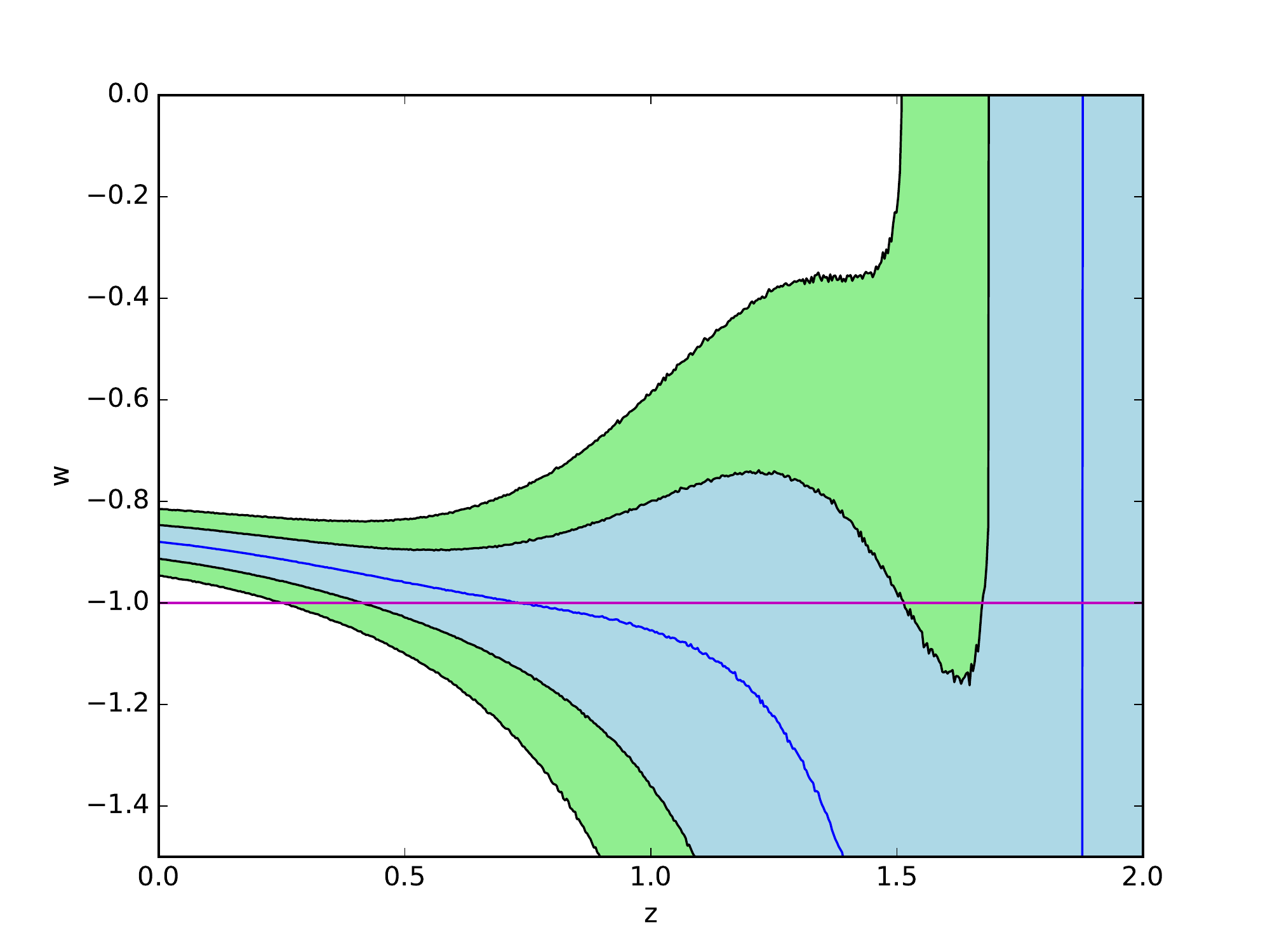}
\includegraphics[scale=0.2]{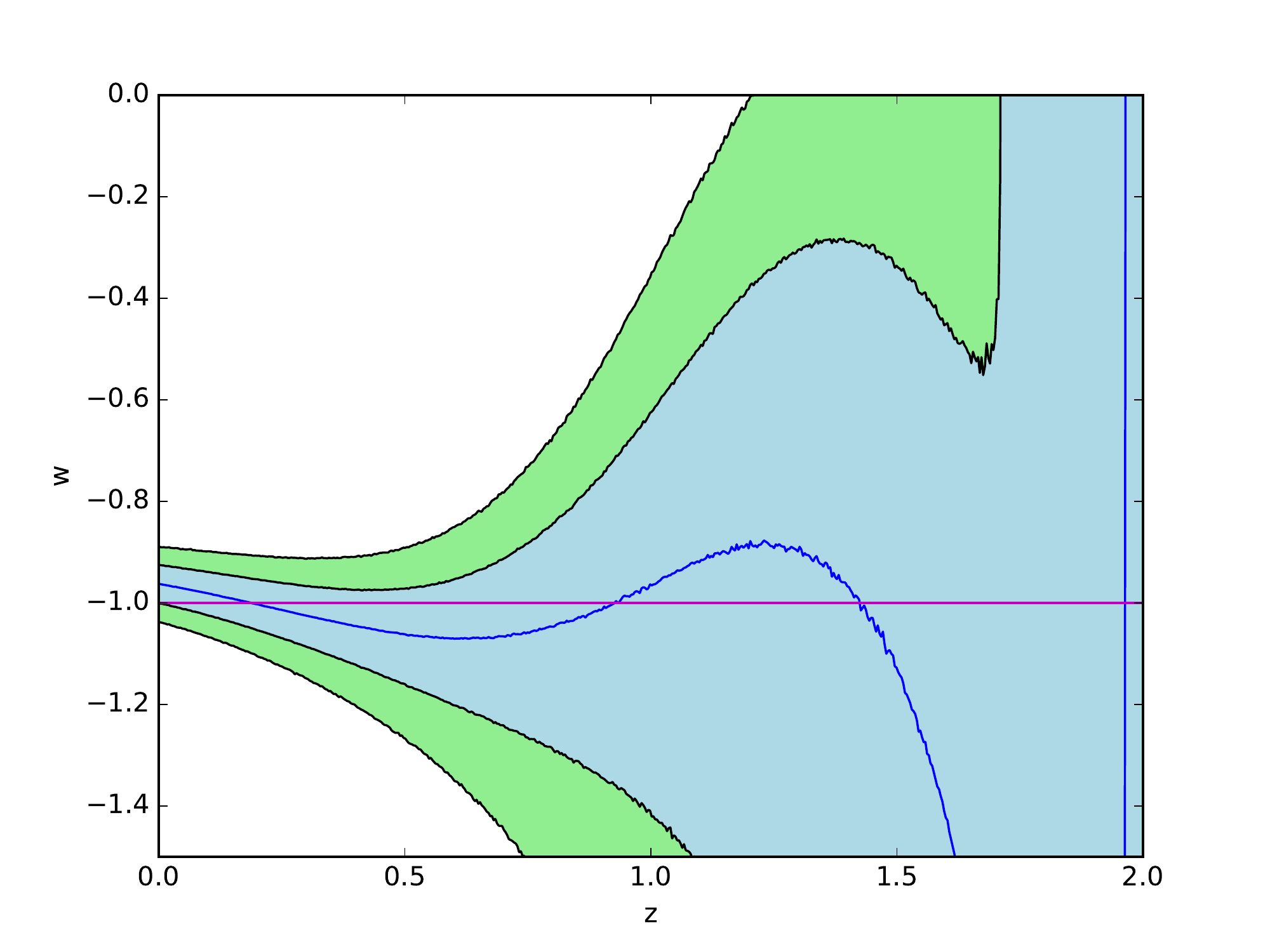}
\caption{The GP reconstructions of the dark energy EoS $\omega(z)$ using SNe Ia + H(z) + CMB. The upper panels from left to right correspond to the cases of $\Omega_{k0}=0.005$, $-0.005$, $0.05$ and $-0.04$, respectively. The lower panels from left to right correspond to the cases of $\Omega_{k0}=0.2$, $-0.2$ and $-0.08$, respectively. We have assumed $\Omega_{m0}=0.308\pm0.012$ and $H_0=73.24\pm1.74$ km s$^{-1}$ Mpc$^{-1}$.}\label{f4}
\end{figure}
\begin{figure}
\centering
\includegraphics[scale=0.4]{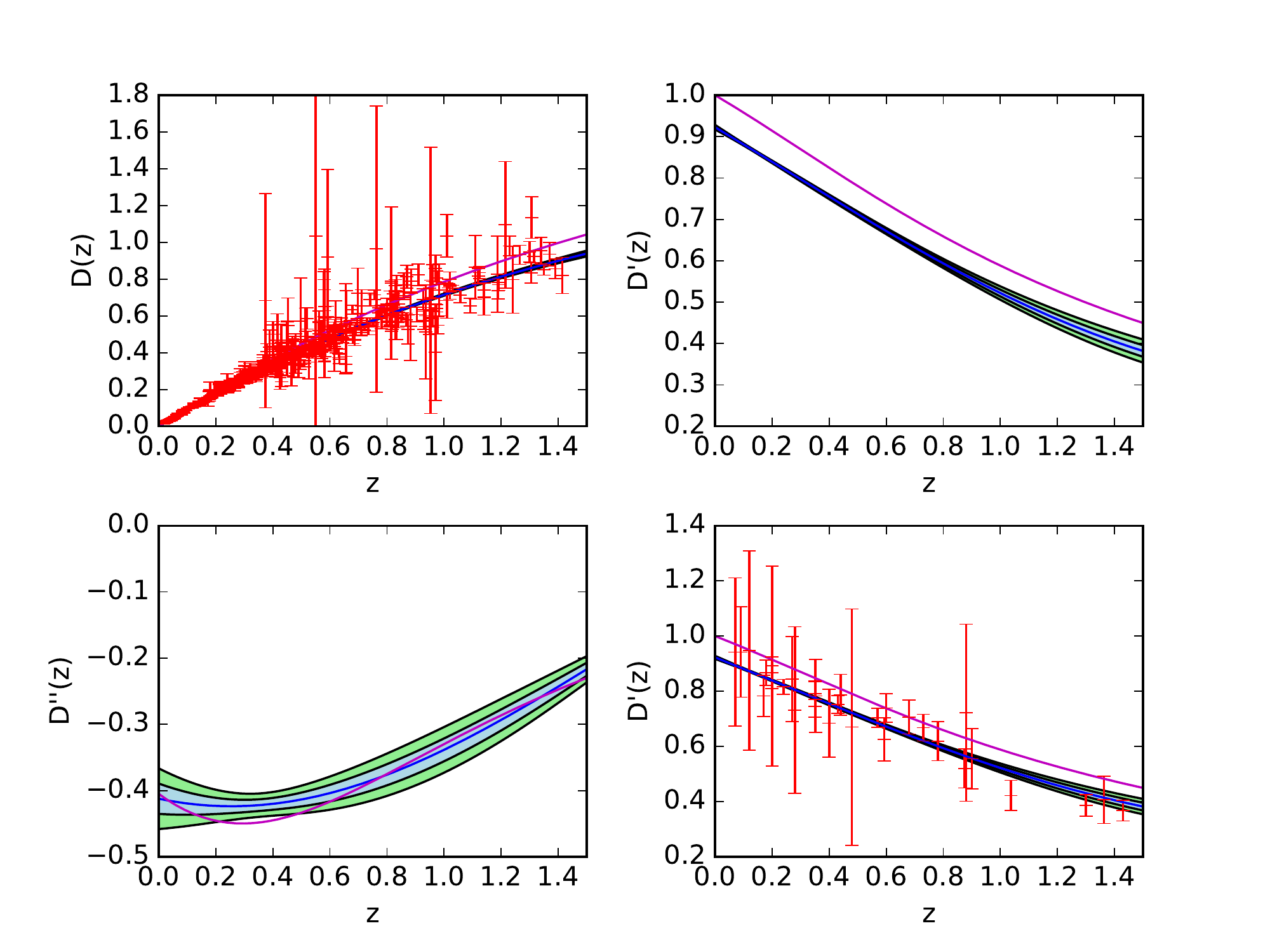}
\includegraphics[scale=0.4]{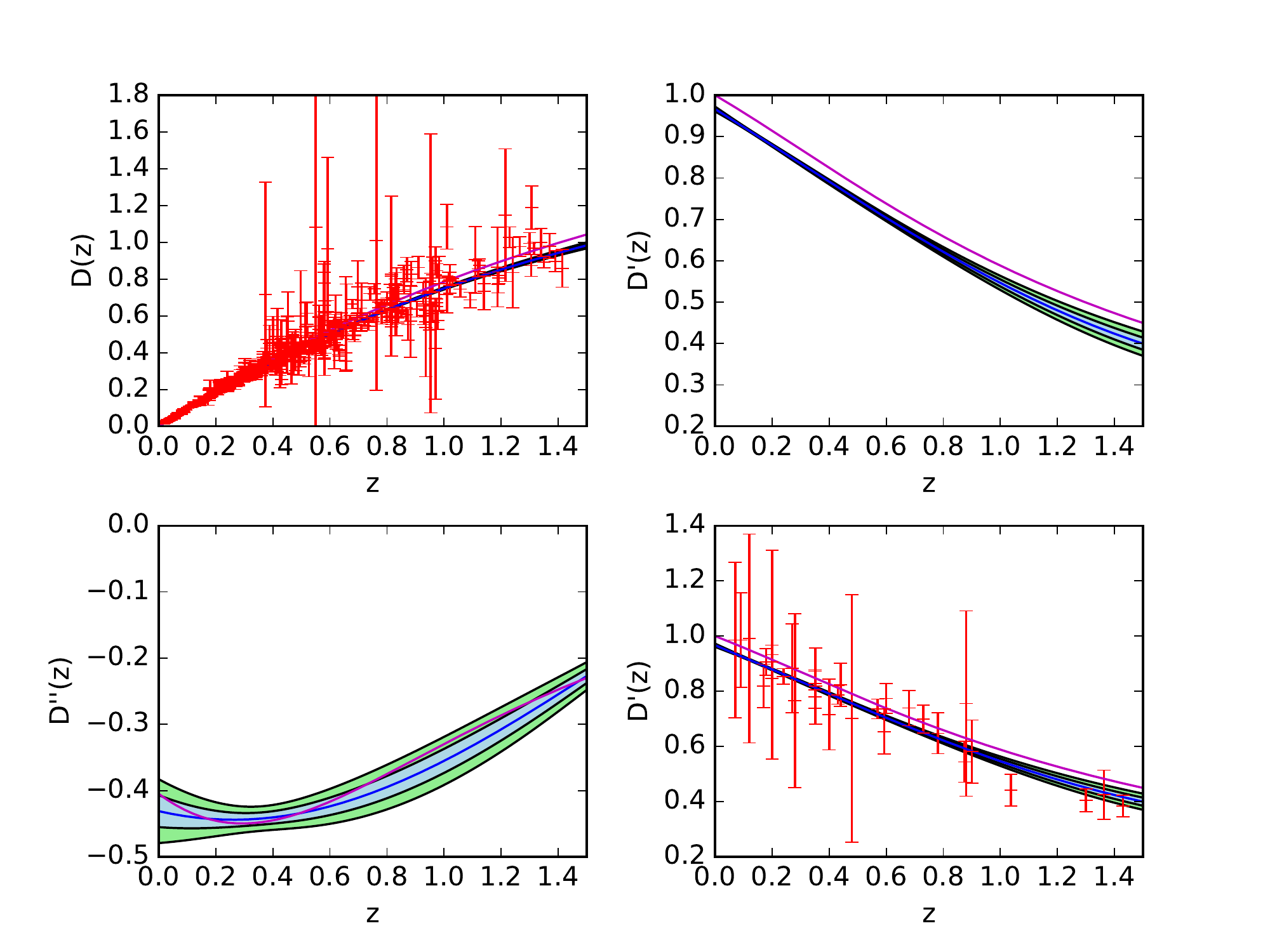}
\includegraphics[scale=0.4]{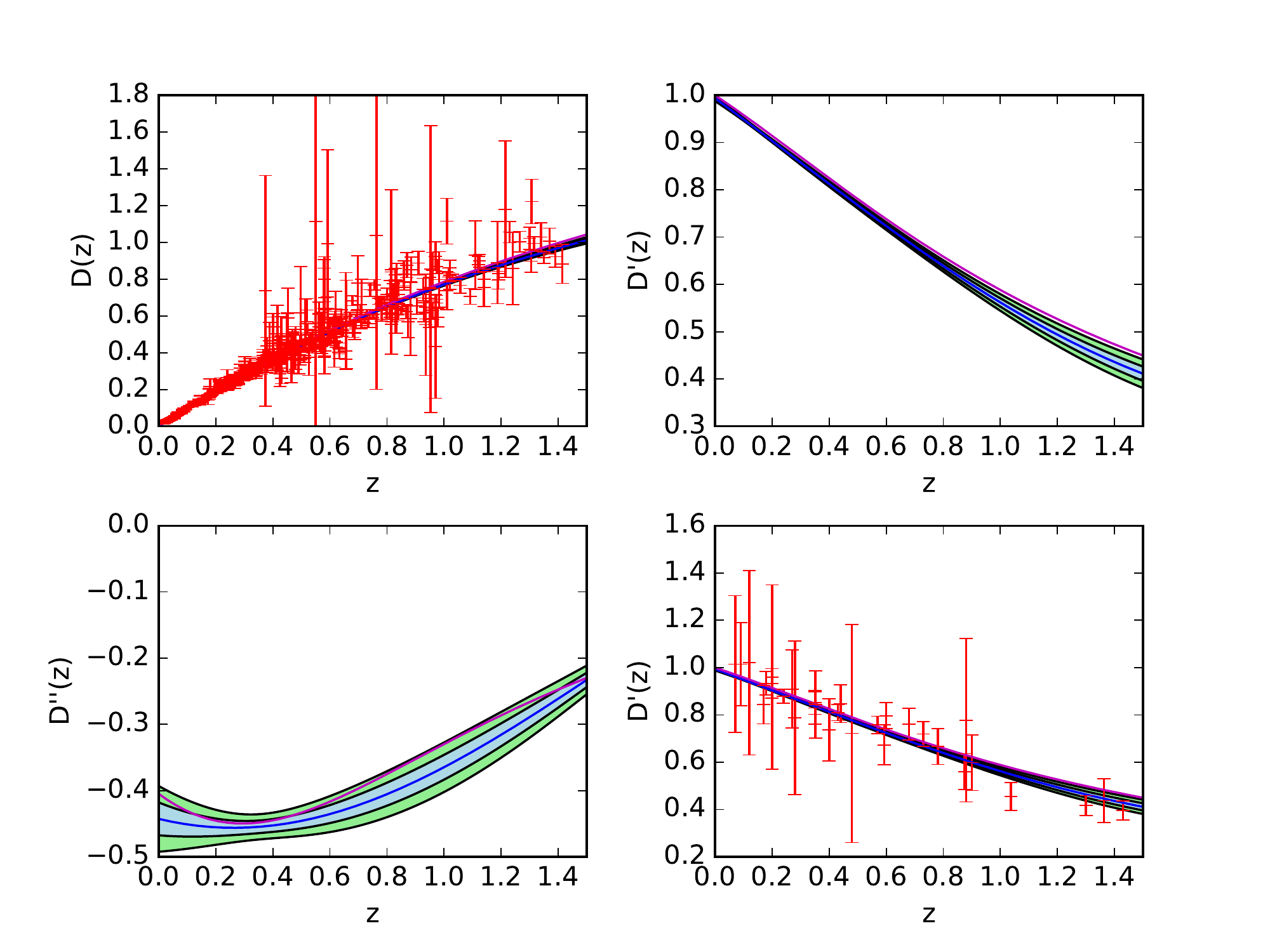}
\includegraphics[scale=0.4]{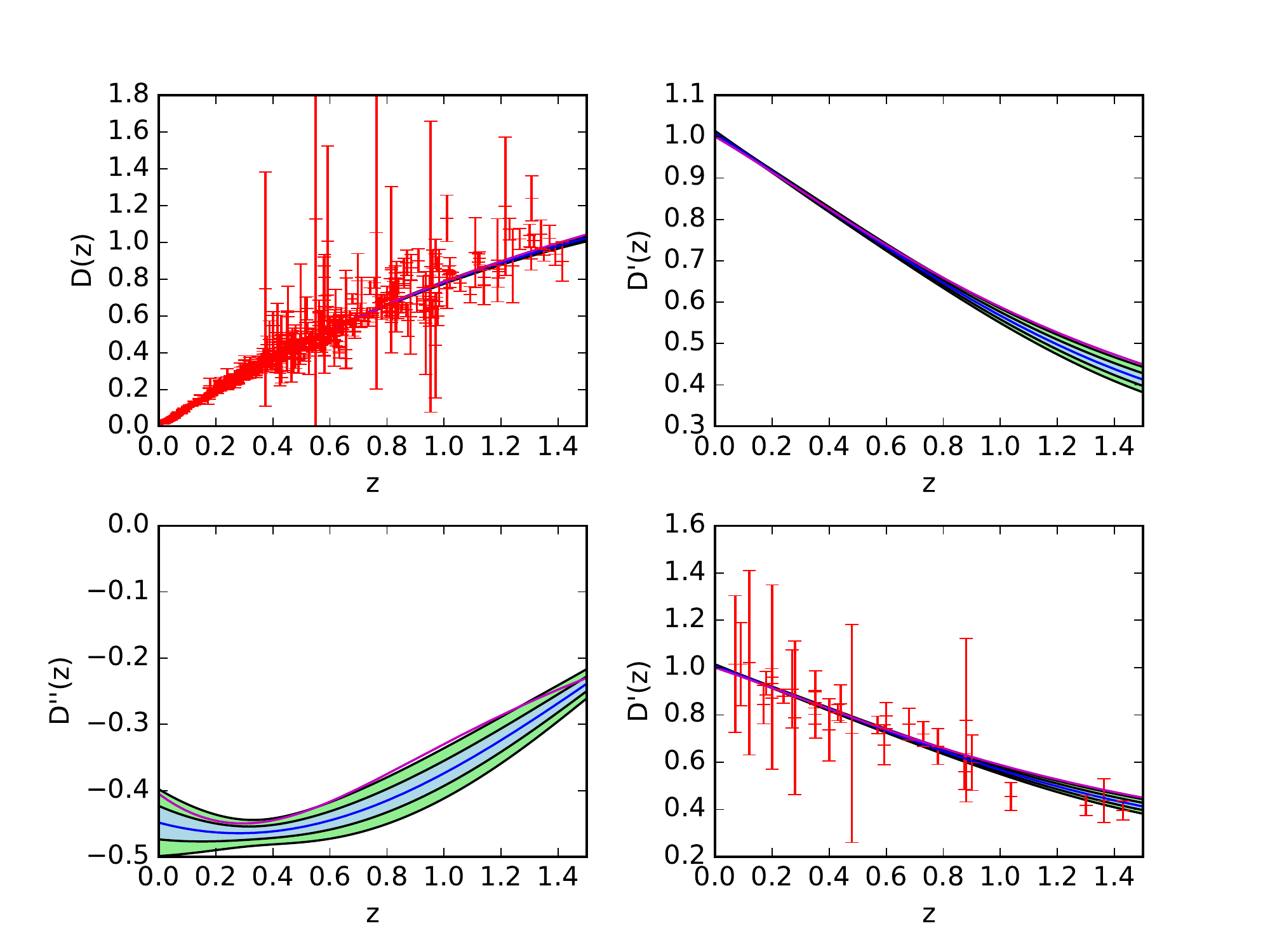}
\includegraphics[scale=0.4]{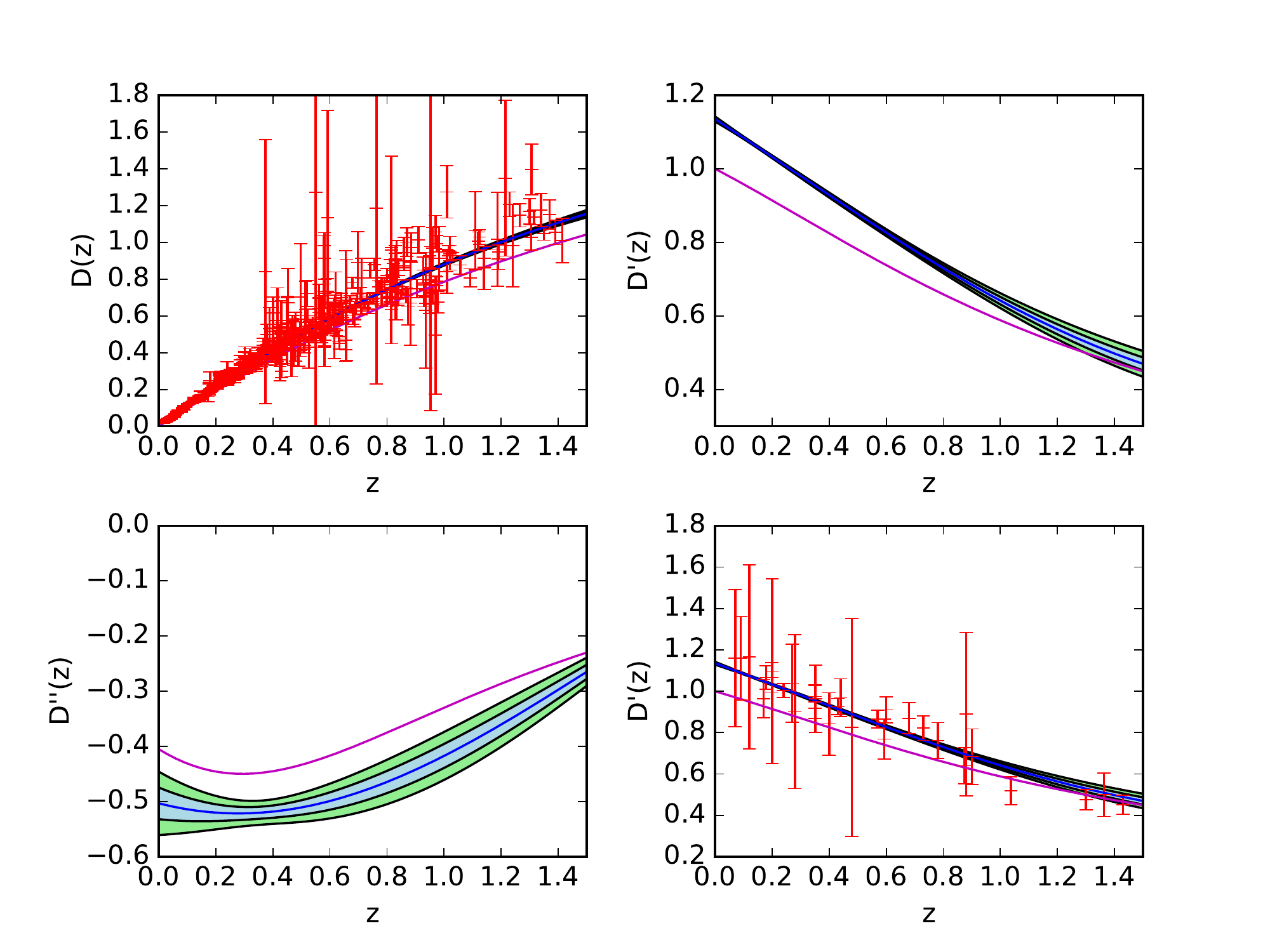}
\includegraphics[scale=0.4]{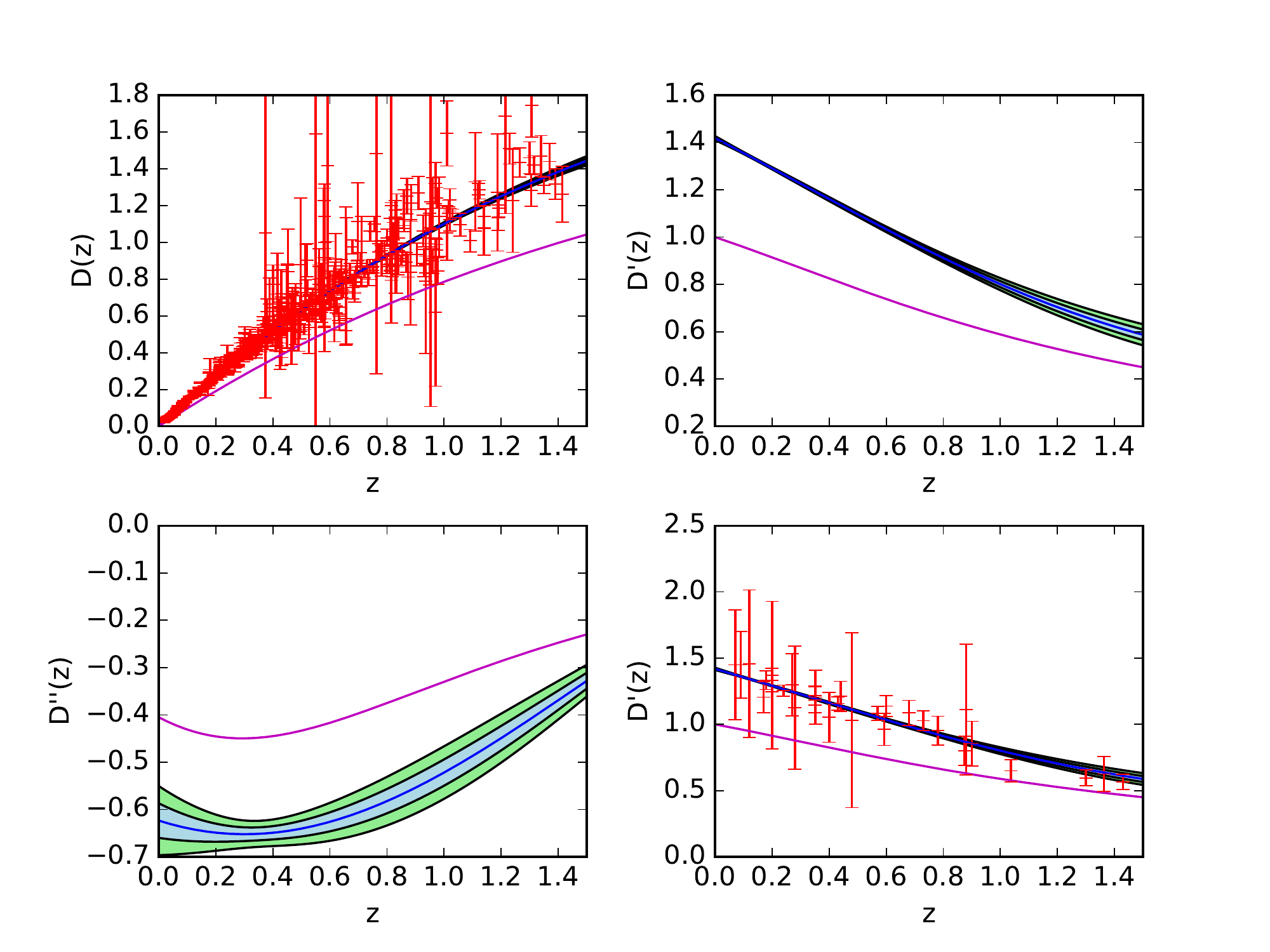}
\caption{The GP reconstructions of $D(z), D'(z)$ and $D''(z)$ using SNe Ia + H(z) + CMB. The upper panels from left to right correspond to the cases of $H_0=65$ and $68.1$ km s$^{-1}$ Mpc$^{-1}$, respectively. The medium panels from left to right correspond to the cases of $H_0=70$ and $73.8$ km s$^{-1}$ Mpc$^{-1}$, respectively. The lower panels from left to right correspond to the cases of $H_0=80$ and $100$ km s$^{-1}$ Mpc$^{-1}$, respectively. We have assumed $\Omega_{m0}=0.308\pm0.012$ and $\Omega_{k0}=0$.}\label{f5}
\end{figure}
\begin{figure}
\centering
\includegraphics[scale=0.23]{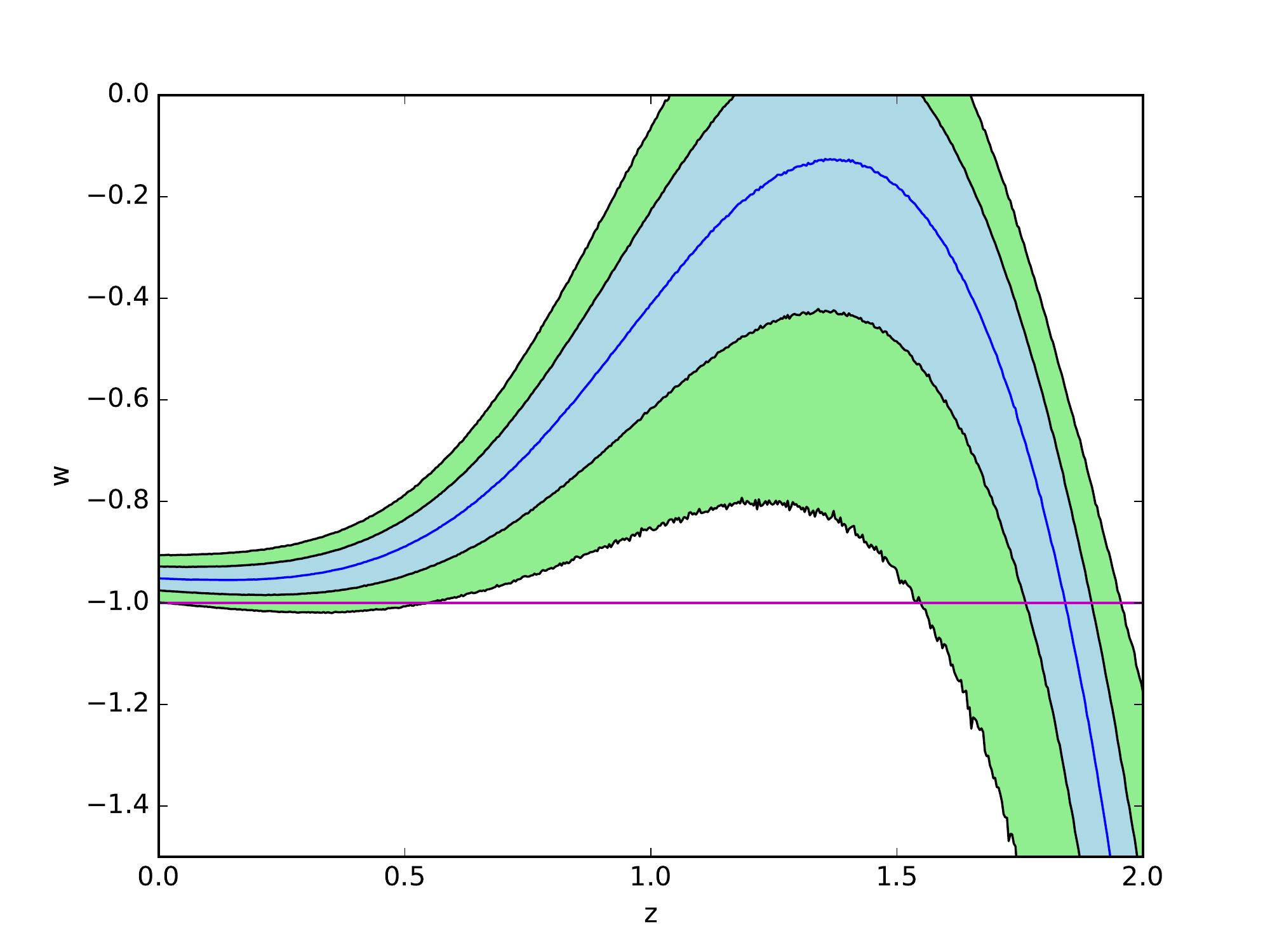}
\includegraphics[scale=0.23]{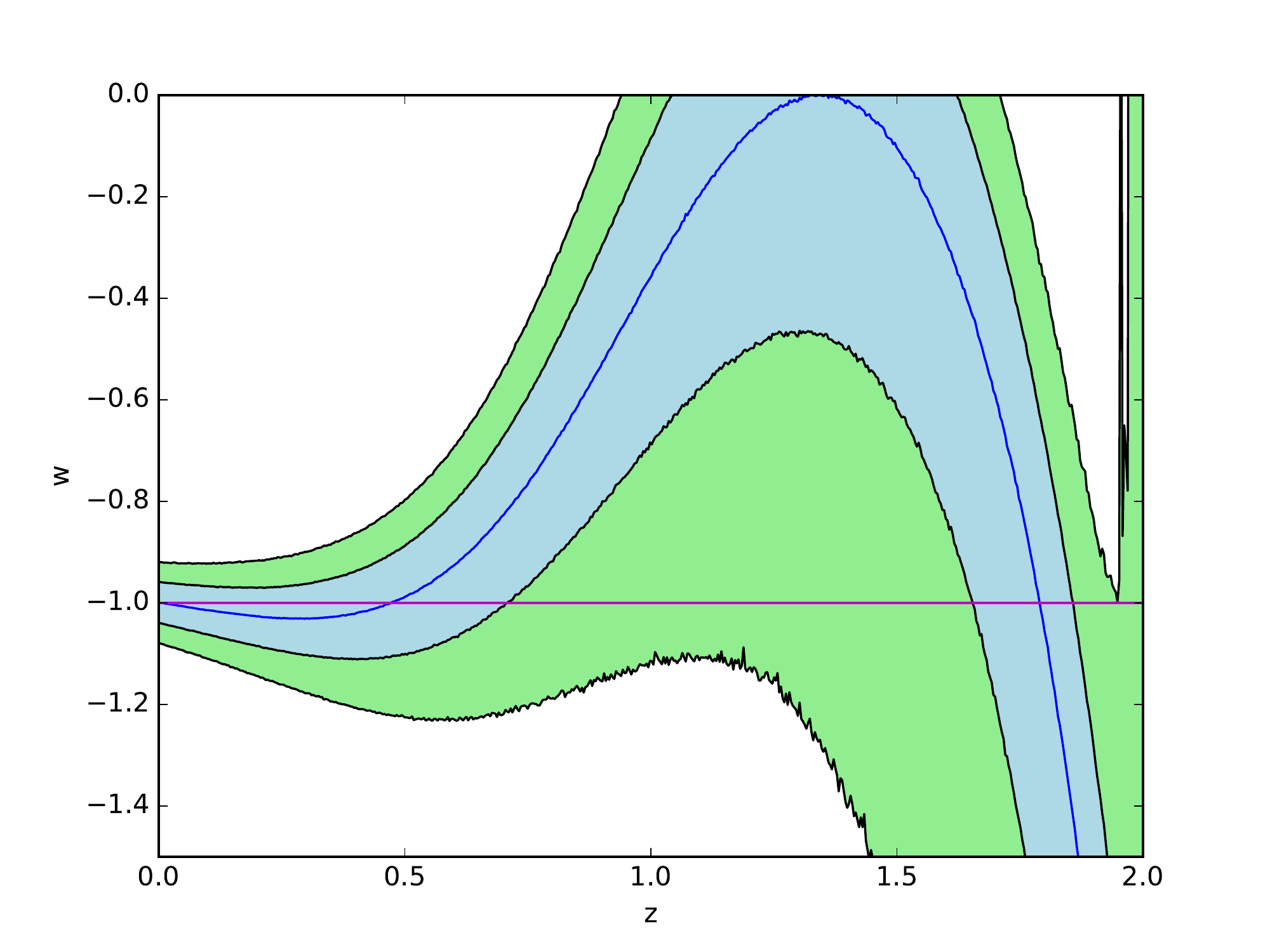}
\includegraphics[scale=0.23]{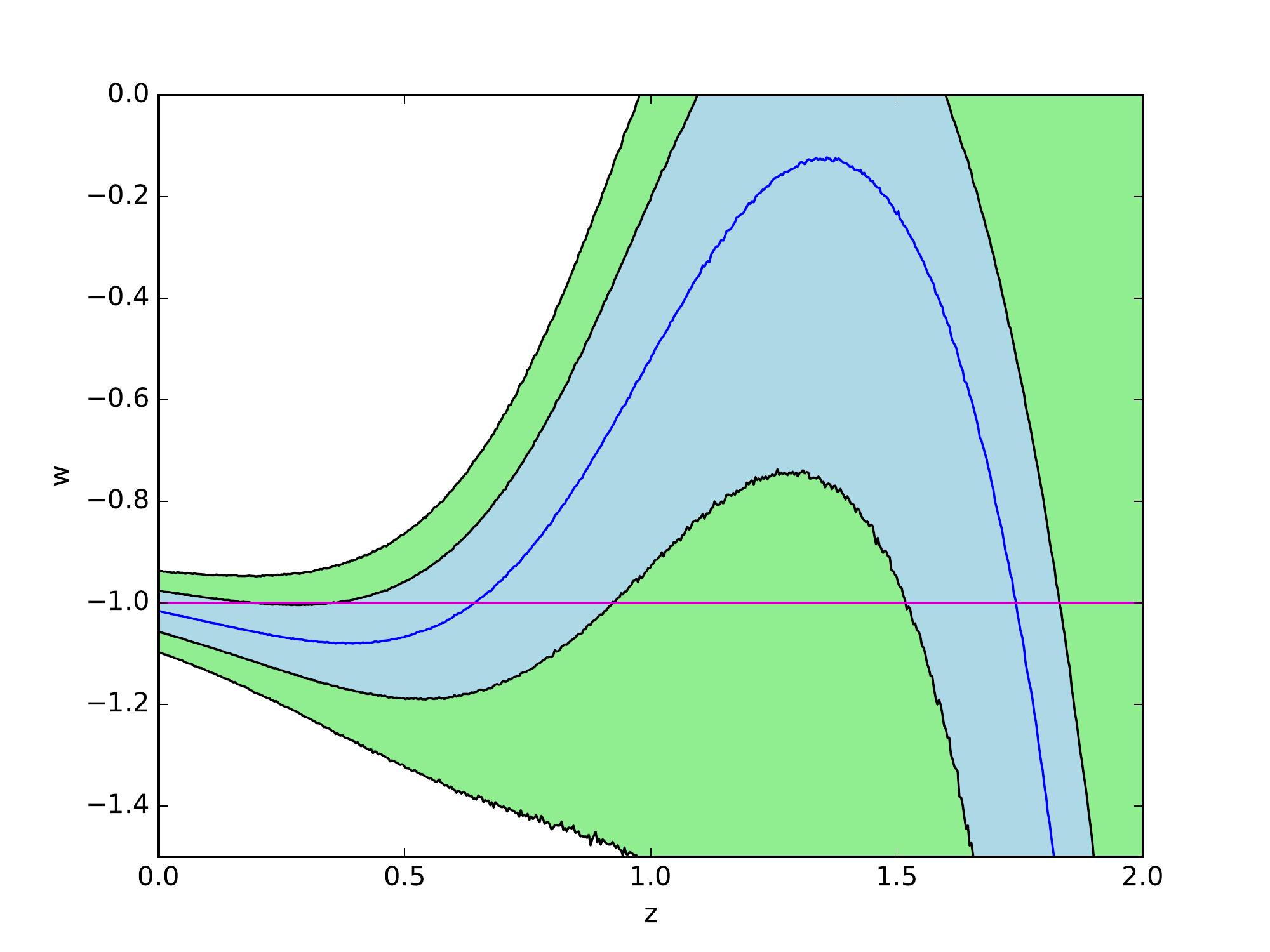}
\includegraphics[scale=0.23]{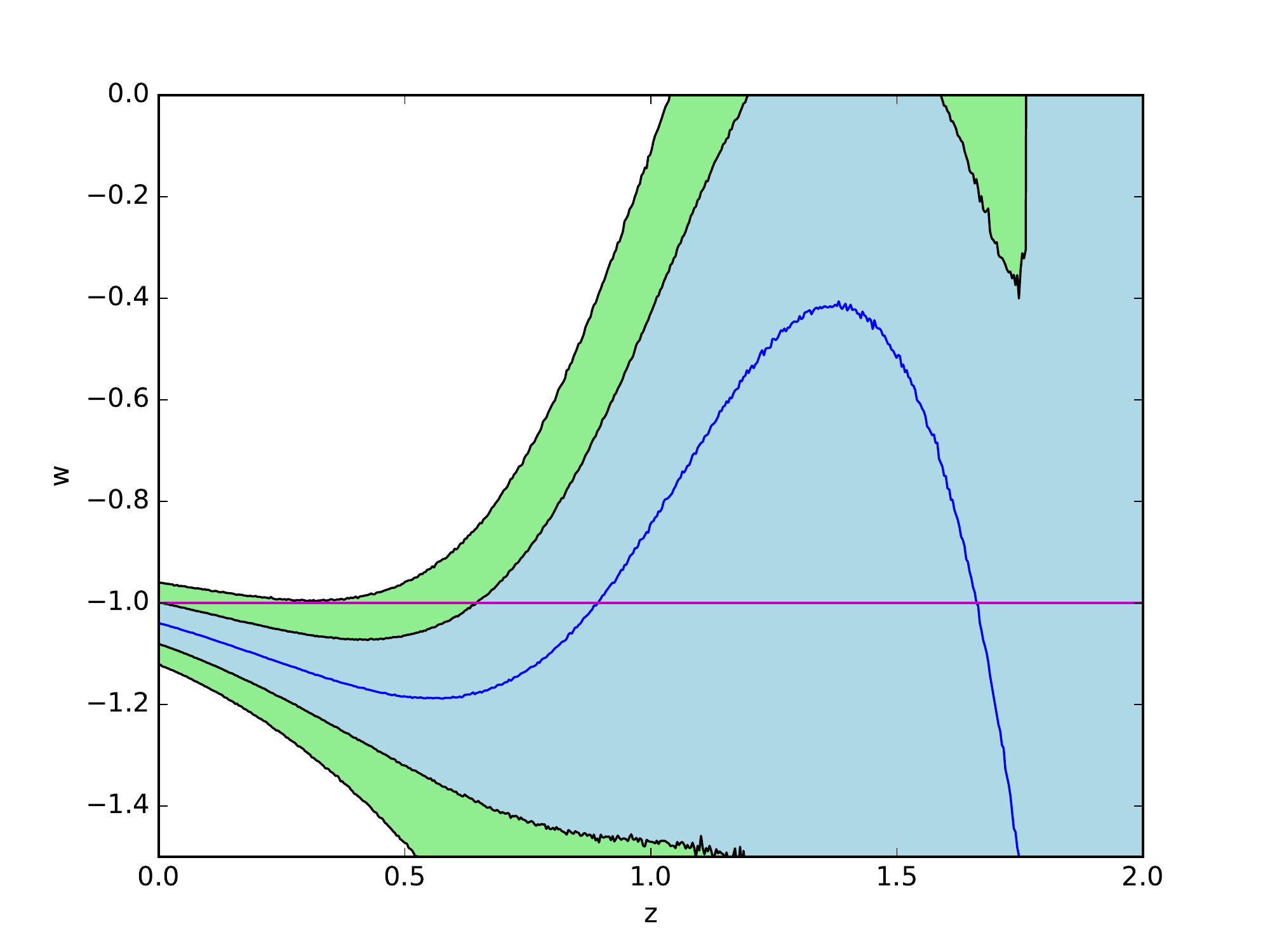}
\includegraphics[scale=0.23]{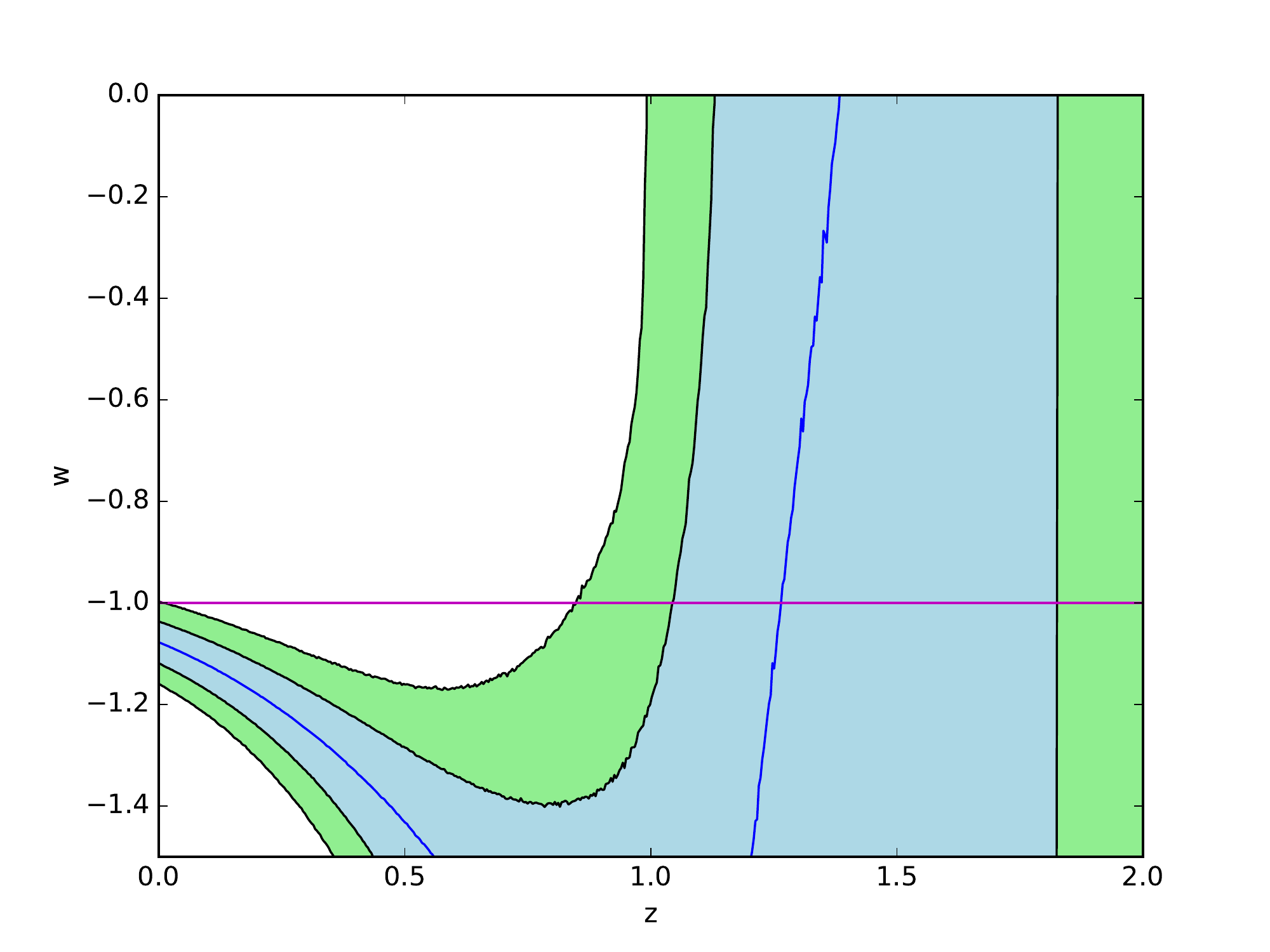}
\includegraphics[scale=0.23]{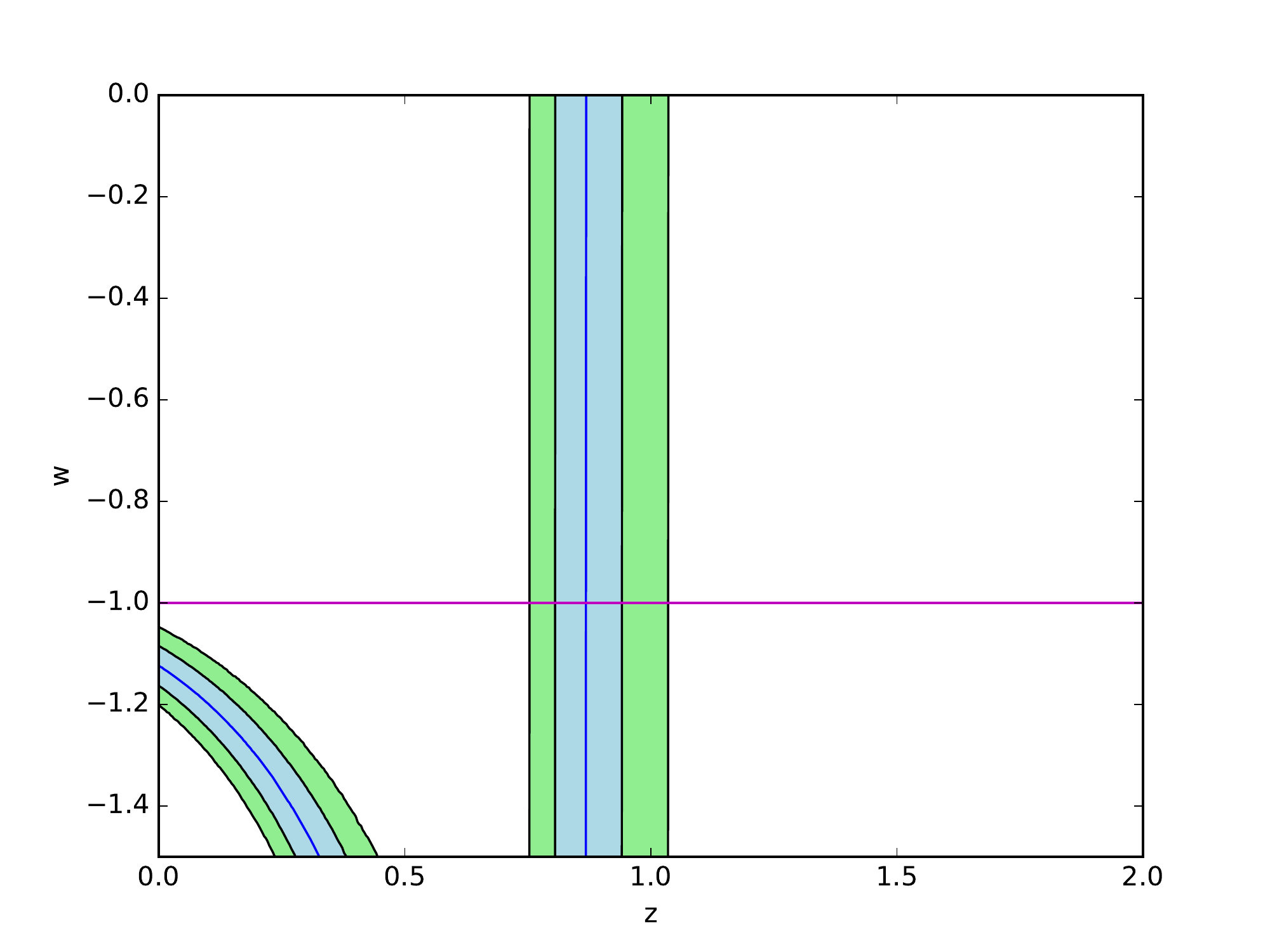}
\caption{The GP reconstructions of the dark energy EoS $\omega(z)$ using SNe Ia + H(z) + CMB. The upper panels from left to right correspond to the cases of $H_0=65$, $68.1$ and $70$ km s$^{-1}$ Mpc$^{-1}$, respectively. The lower panels from left to right correspond to the cases of $H_0=73.8$, $80$ and $100$ km s$^{-1}$ Mpc$^{-1}$, respectively. We have assumed $\Omega_{m0}=0.308\pm0.012$ and $\Omega_{k0}=0$.}\label{f6}
\end{figure}
\begin{figure}
\centering
\includegraphics[scale=0.38]{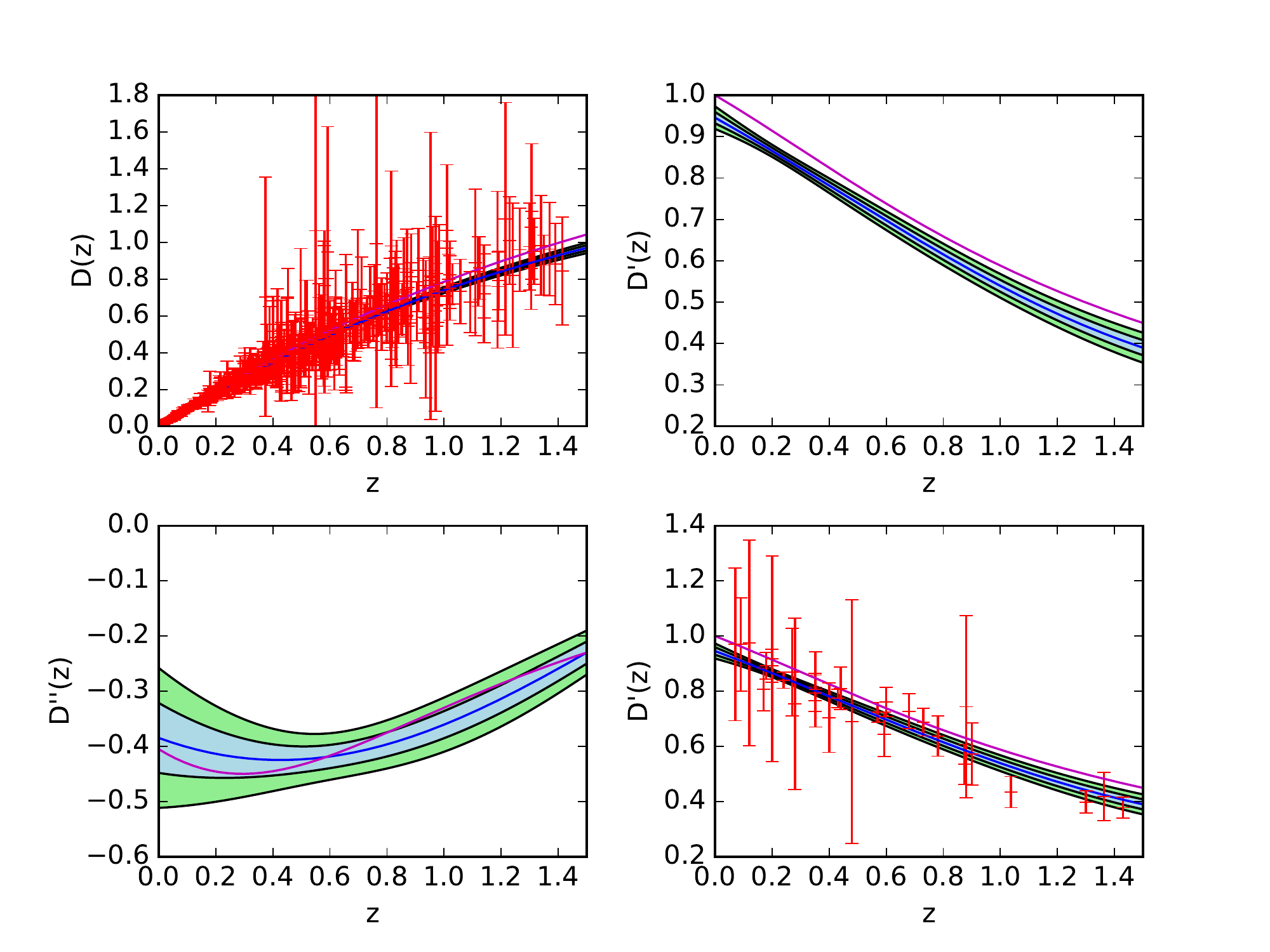}
\includegraphics[scale=0.38]{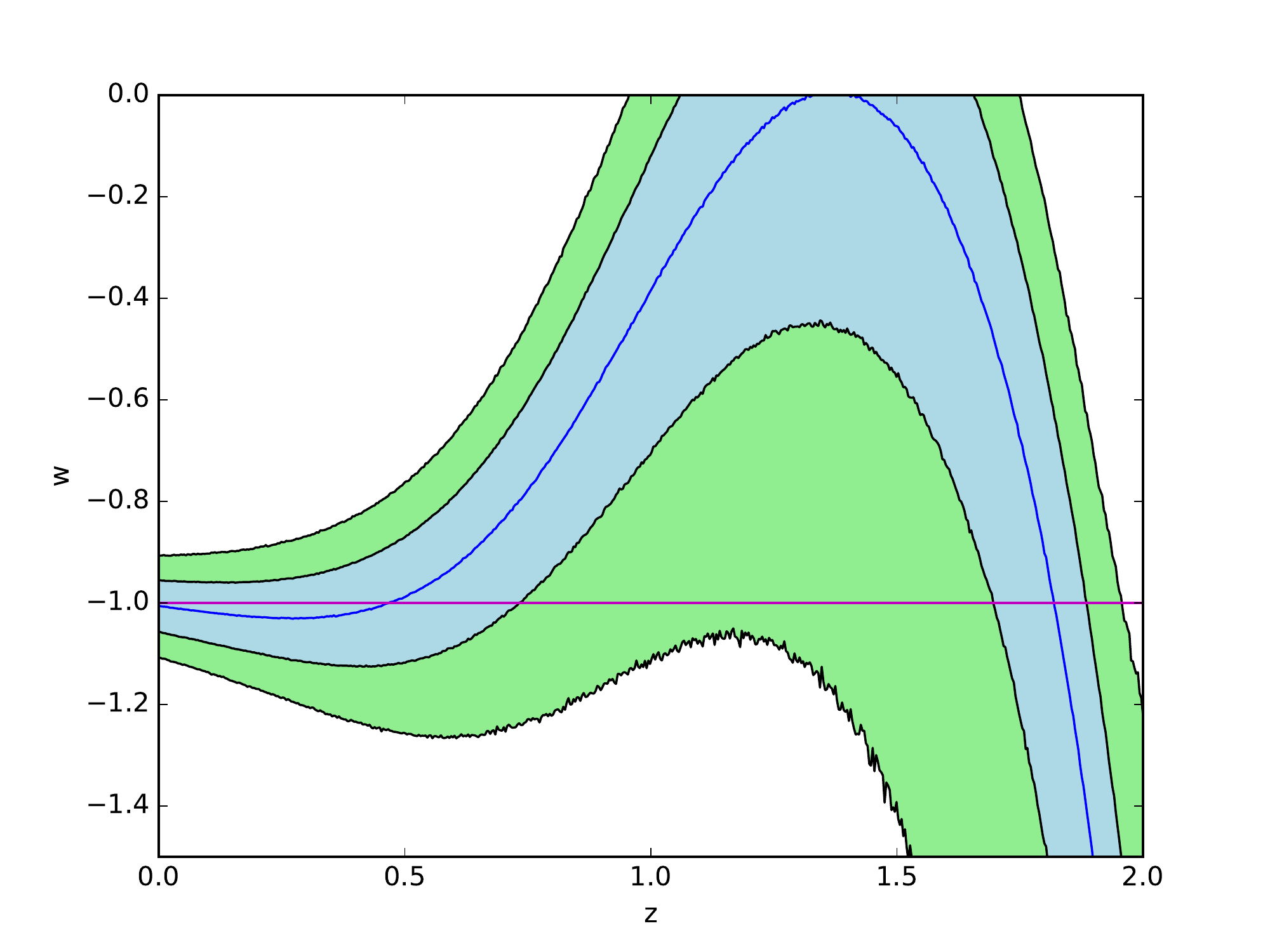}
\caption{The GP reconstructions of $D(z), D'(z)$, $D''(z)$ and the dark energy EoS $\omega(z)$ using SNe Ia + H(z) + CMB. We have assumed $\Omega_{m0}=0.308\pm0.012$, $\Omega_{k0}=0$ and $H_0=66.93\pm0.62$ km s$^{-1}$ Mpc$^{-1}$.}\label{f7}
\end{figure}

In the upper left panel of Fig. \ref{f4}, one can find that the $\Lambda$CDM model happens to lie on the $2\sigma$ confidence region when $\Omega_{k0}=0.005$. When adopting a negative value $\Omega_{k0}=-0.005$, the $\Lambda$CDM model still lies in the $2\sigma$ confidence region. Subsequently, increasing an order of magnitude, one can find that the $\Lambda$CDM model lies out the $2\sigma$ confidence region in the low-redshift range when $\Omega_{k0}=0.05$, and happens to lie on the $1\sigma$ confidence region when $\Omega_{k0}=-0.04$ (see the two upper right panel of Fig. \ref{f4}). To determine the possible range of $\Omega_{k0}$, we also consider the cases of $\Omega_{k0}=0.2$, $-0.2$ and $-0.08$, respectively (see the lower panels of Fig. \ref{f4}). When $\Omega_{k0}=0.2$, the $\Lambda$CDM model lies out the $2\sigma$ confidence region in the low-redshift range very apparently. Conversely, when $\Omega_{k0}=-0.2$, the same deviation also occurs in the low-redshift range. However, when $\Omega_{k0}=-0.08$, the $\Lambda$CDM model happens to lie on the $1\sigma$ confidence region. Therefore, our GP reconstruction can provide a relatively tight constraint on $\Omega_{k0}$ at $1\sigma$ confidence level, i.e., $\Omega_{k0}\in[-0.08,-0.04]$, which is an order of magnitude higher than the recent Planck's predictions $\Omega_{k0}<|0.005|$. In the meanwhile, the effects of variable $\Omega_{k0}$ on the reconstructions of $D(z), D'(z)$ and $D''(z)$ is too small to consider them. From the point of view of observational data, $\Omega_{k0}$ is not associated with SNe Ia and $H(z)$ data, and is only included in the Planck's shift parameter through the dimensionless Hubble parameter $E(z)$. As a result, its effects on the normalized comoving distance can be ignored.

It is worth noting that in the previous literature, as the case of $\Omega_{m0}$, the authors can still not give out a relatively tight constraint on $\Omega_{k0}$ since they do not utilize the $H(z)$ and CMB data.

\subsection{The effects of variable $H_0$ on the dark energy EoS}
From Eq. (\ref{4}) and the above-mentioned `` $\textit{relations}$ '', one can conclude that the Hubble constant $H_0$ is directly associated with SNe Ia and $H(z)$ data. Thus, its different values will affect the reconstructions of both the normalized comoving distance and dark energy EoS. In this subsection, we would like to study the effects of $H_0$ on both the normalized comoving distance and dark energy EoS by assuming $\Omega_{m0}=0.308\pm0.012$ and $\Omega_{k0}=0$.

At first, in the upper panels of Fig. \ref{f5}, we consider two relatively small cases of $H_0$, i.e., $65$ and $68.1$ km s$^{-1}$ Mpc$^{-1}$. It is easy to see that the $\Lambda$CDM model lies out the $2\sigma$ confidence region obviously in the relatively high-redshift range for the reconstructions of $D$ and $D'$ of these two cases. This indicates that too small Hubble constant is disfavored by our GP reconstructions based on currently cosmological observations. In the medium panels of Fig. \ref{f5}, we consider two relatively large cases, i.e., $70$ and $73.8$ km s$^{-1}$ Mpc$^{-1}$. One can easily conclude that the $\Lambda$CDM model lies well in the $2\sigma$ confidence region for these two cases, especially for the latter case. However, the best constraint is still our standard case (see the upper left panel of Fig. \ref{f3}). Furthermore, since the latest local value of $H_0$ measured by Riess et al. is higher than our expectation, we predict boldly that the value of $H_0$ may continue to increase in the future. Hence, it is substantially necessary and constructive to investigate the effects of larger $H_0$ on the reconstructions of $D(z), D'(z)$ and $D''(z)$. Based on this concern, in the lower panels of Fig. \ref{f5}, we consider two relatively large cases of $H_0$, i.e., $80$ and $100$ km s$^{-1}$ Mpc$^{-1}$. One can find that these two cases deviate apparently from the $H(z)$ data. Especially, for the latter case, the $\Lambda$CDM model is very inconsistent with the predictions of our GP reconstructions over $2\sigma$ level. This indicates that too large Hubble constant is disfavored by currently cosmological observations, and that the measurement of the Hubble constant plays an extremely important role in modern precise cosmology.

In what follows, we would like to investigate the effects of $H_0$ on the dark energy EoS. It is very clear that $H_0$ affects the reconstructions of the dark energy EoS by affecting those of $D(z), D'(z)$ and $D''(z)$.

In the upper left panel of Fig. \ref{f6}, the $\Lambda$CDM model lies out the $2\sigma$ confidence region in the low-redshift range when $H_0=65$ km s$^{-1}$ Mpc$^{-1}$. This implies that too small Hubble constant is disfavored by our GP reconstructions based on current cosmological observations. In the upper medium panel of Fig. \ref{f6}, one can find that the $\Lambda$CDM model happens to lie on the $2\sigma$ confidence region when $H_0=68.1$ km s$^{-1}$ Mpc$^{-1}$. Subsequently, as done in \cite{6}, we also consider the case of $H_0=70$ km s$^{-1}$ Mpc$^{-1}$, and find that the $\Lambda$CDM model is constrained more strictly and lies in the $2\sigma$ confidence region. In the lower left panel of Fig. \ref{f6}, when $H_0=73.8$ km s$^{-1}$ Mpc$^{-1}$, the $\Lambda$CDM model happens to lie on the $2\sigma$ confidence region. Thus, our GP reconstructions can provide a relatively tight constraint on $H_0$ at $2\sigma$ confidence level, i.e., $H_0\in[68.1,73.8]$ km s$^{-1}$ Mpc$^{-1}$. Furthermore, as before, we also take into account the effects of larger $H_0$ on the dark energy EoS. We find that from the two lower right panels of Fig. \ref{f6}, the larger $H_0$ is, the more apparently our GP reconstructions deviate from the $\Lambda$CDM model.

Notice that the relatively tight constraint on the Hubble constant $H_0$ is still ascribed to the newly added $H(z)$ and CMB data.

\subsection{The $H_0$ tension}
Recently, the $H_0$ tension reported by Riess et al. has attracted a lot of attention. We are also very interested in exploring it by utilizing GP reconstructions. More concretely, assuming $\Omega_{m0}=0.308\pm0.012$, $\Omega_{k0}=0$ and $H_0=66.93\pm0.62$, we find that $\Lambda$CDM model lies out the $2\sigma$ confidence region for the reconstructions of $D$ and $D'$, and is not compatible with the correspondingly reconstructed dark energy EoS over $2\sigma$ level when $z>1.95$ (see Fig. \ref{f7}). Interestingly, comparing this with the case of $\Omega_{m0}=0.308\pm0.012$, $\Omega_{k0}=0$ and $73.24\pm1.74$ (see Fig. \ref{f2} and the upper left panel of Fig. \ref{f3}), we find that the results of our reconstructions support substantially the recent local measurement of $H_0$ reported by Riess et al.

\section{Discussions and conclusions}
One of the most important problems in modern cosmology is to determine whether the dark energy is the cosmological constant. We are motivated by using the model-independent GP method to exhibit a comprehensive investigations of the dark energy EoS. Different from the previous literature, we utilize the `` controlling variable method '' to study directly the effects of variable matter density parameter $\Omega_{m0}$, variable cosmic curvature $\Omega_{k0}$ and  variable Hubble constant $H_0$ on the dark energy EoS, respectively.

First of all, we modify the usual GaPP code by using the newly added $H(z)$ data and Planck's shift parameter. It is very obvious that, using a combination of the Union 2.1 data set, $H(z)$ data and Planck's shift parameter, our reconstructions of $D(z), D'(z)$ and $D''(z)$ give out a tighter constraint than those in Fig. 8 of \cite{6}, and is approximately consistent with the $\Lambda$CDM model at $1\sigma$ level (see Fig. \ref{f2}). To be more precise, the newly added H(z) data gives out a stricter constraint in the low-redshift range, and the Planck's shift parameter gives out a tighter high-redshift constraint avoiding the divergence in the high-redshift range. Therefore, we can reconstruct the dark energy EoS better using the stricter constraint. By comparing the effects of different probe on the reconstructions of $D(z), D'(z)$, $D''(z)$ and the dark energy EoS, we find that the $H(z)$ probe plays the main role in improved constraints on the dark energy EoS.

In the second place, assuming $\Omega_{k0}=0$ and $H_0=73.24\pm1.74$ km s$^{-1}$ Mpc$^{-1}$, we find that too small or large $\Omega_{m0}$ can be directly ruled out using currently cosmological observations. In the meanwhile, making full use of our modified code, we find that our GP reconstruction provides a relatively tight constraint on $\Omega_{m0}$ at $2\sigma$ confidence level, i.e., $\Omega_{m0}\in[0.268,0.310]$. Subsequently, assuming $\Omega_{m0}=0.308\pm0.012$ and $H_0=73.24\pm1.74$ km s$^{-1}$ Mpc$^{-1}$, we also find that too small or large $\Omega_{k0}$ is disfavored by our GP reconstructions based on currently cosmological observations. At the same time, our GP reconstruction also provides a relatively tight constraint on $\Omega_{k0}$ at $1\sigma$ confidence level, i.e., $\Omega_{k0}\in[-0.08,-0.04]$, which is an order of magnitude higher than the recent Planck's predictions $\Omega_{k0}<|0.005|$. It is noteworthy that the effects of variable $\Omega_{m0}$ and $\Omega_{k0}$ on the reconstructions of $D(z), D'(z)$ and $D''(z)$ are too small to take into account them. From the point of view of observational data, $\Omega_{m0}$ and $\Omega_{k0}$ are only included in the Planck's shift parameter and not associated with SNe Ia and $H(z)$ data. Consequently, their effects on the normalized comoving distance can be ignored. Furthermore, assuming $\Omega_{m0}=0.308\pm0.012$ and $\Omega_{k0}=0$, we also find that too small or large $H_0$ is disfavored by our GP reconstructions using currently cosmological observations, and that our GP reconstructions provide a relatively tight constraint on $H_0$ at $2\sigma$ confidence level, i.e., $H_0\in[68.1,73.8]$ km s$^{-1}$ Mpc$^{-1}$. Different from the cases of $\Omega_{m0}$ and $\Omega_{k0}$, variable $H_0$ affects apparently the reconstructions of the dark energy EoS by affecting those of $D(z), D'(z)$ and $D''(z)$. This indicates that the measurement of the Hubble constant plays an extremely important role in modern cosmology.

In addition, assuming $\Omega_{m0}=0.308\pm0.012$ and $\Omega_{k0}=0$, we find that the results of our GP reconstructions support the recent local measurement of $H_0$ reported by Riess et al. very much.

Note that the best constraint that we can provide indicates that the GP reconstructions are still consistent with $\Lambda$CDM model at $2\sigma$ level (see Fig. \ref{f2} and the upper left panel of Fig. \ref{f3}). Therefore, we expect more high-quality SNe Ia, $H(z)$ and CMB data can give stricter constraint on the dark energy EoS in the future.

\section{acknowledgements}
This study is supported in part by the National Science Foundation of China. The author Xin-He Meng thanks Professors S. D. Odintsov and Bharat Ratra for beneficial discussions on cosmology. The author Deng Wang warmly thanks Qi-Xiang Zou for helpful communications and programming for a long time.

\end{document}